\documentclass[twocolumn]{revtex4-1}
\usepackage{amssymb}
\usepackage{dcolumn}
\usepackage{bm}
\usepackage{amsfonts}
\usepackage{amsmath}
\usepackage{subfigure}
\usepackage{graphicx}

\newtheorem{theorem}{Theorem}

\begin{document}

\title{A microscopic two--band model for the electron-hole
asymmetry in high-$T_c$ superconductors and reentering behavior}
\author{J.-B. Bru, W. de Siqueira Pedra and A.-S. D{\"o}mel \vspace{1.5cm}}

\date{\today}

\begin{abstract}
To our knowledge there is no rigorously analyzed microscopic model explaining the electron--hole asymmetry of the
critical temperature seen in high--$T_{c}$ cuprate superconductors -- at least no model not breaking artificially this symmetry.
We present here a microscopic two--band model based on the structure of energetic levels of holes in
$\mathrm{CuO}_{2}$ conducting layers of cuprates.  In particular,
our Hamiltonian does not contain \emph{ad hoc} terms implying -- explicitly -- different masses for electrons and holes.
We prove that two energe\-tically near--lying interacting bands can explain
 the electron--hole asymmetry.
Indeed, we  rigorously
analyze the phase diagram of the model and show that
 the critical temperatures
 for fermion densities below half--filling can manifest a very different behavior as compared to the case of densities above half--filling. This fact results from the inter--band interaction
 and intra--band Coulomb repulsion in interplay with thermal fluctuations between two energetic levels. So, if the energy difference between
bands is too big (as compared to the energy scale defined by the critical temperatures of
 superconductivity) then the asymmetry disappears.
Moreover, the critical temperature turns out to be a non--monotonic function
of the fermion density and the phase diagram of our model shows
``superconducting domes'' as in high--$T_{c}$ cuprate
superconductors. This explains why
the maximal critical temperature is attained at donor densities away from the
maximal one. Outside the superconducting phase and for fermion densities
near half--filling the thermodynamics
 governed by our Hamiltonian corresponds,  as in real high--$T_c$ materials,
to a Mott--insulating phase.
The nature of the inter--band interaction can be
 electrostatic (screened Coulomb interaction), magnetic
(for instance some Heisenberg--type one--site spin--spin interaction),
 or a mixture of both.
If the inter--band interaction is predominately
magnetic then -- additionally to the electron--hole asymmetry -- we observe a reentering behavior meaning that the superconducting phase can only occur in a finite interval
of temperatures. This phenomenon is rather rare, but has also been observed in the so--called magnetic superconductors.
The mathematical results here
 are direct consequences of [J.-B. Bru, W. de Siqueira Pedra, Rev.
Math. Phys. 22, 233--303 (2010)] which is reviewed in the introduction.
\end{abstract}

\maketitle

\section{Introduction}

Theoretical foundations of superconductivity go back to the celebrated BCS
theory -- appeared in the late fifties (1957) -- which explains conventional
type I superconductors. The lattice version of this theory is based on the
so--called (reduced) BCS\ Hamiltonian%
\begin{eqnarray}
\mathrm{H}_{\Lambda }^{BCS} &:&=\sum\limits_{k\in \Lambda _{N}^{\ast
}}\left( \varepsilon _{k}-\mu \right) \left( \tilde{a}_{k,\uparrow }^{\ast }%
\tilde{a}_{k,\uparrow }+\tilde{a}_{k,\downarrow }^{\ast }\tilde{a}%
_{k,\downarrow }\right)  \notag \\
&&-\frac{1}{N}\sum_{k,q\in \Lambda _{N}^{\ast }}\gamma _{k,q}\tilde{a}%
_{k,\uparrow }^{\ast }\tilde{a}_{-k,\downarrow }^{\ast }\tilde{a}%
_{q,\downarrow }\tilde{a}_{-q,\uparrow }  \label{BCS Hamilt}
\end{eqnarray}%
defined in a cubic box $\Lambda _{N}:=\{\mathbb{Z}\cap \lbrack -L,L]\}^{3}$
of volume $|\Lambda _{N}|=N\geq 2$. We choose without loss of generality $N$
such that $L:=(N^{1/3}-1)/2\in \mathbb{N}$. Here $\Lambda _{N}^{\ast }$ is
the reciprocal lattice of quasi--momenta (periodic boundary conditions) and
the operator $\tilde{a}_{k,\mathrm{s}}^{\ast }$ (resp. $\tilde{a}_{k,\mathrm{%
s}}$) creates (resp. annihilates) a fermion with spin $\mathrm{s}\in
\{\uparrow ,\downarrow \}$ and (quasi--) momentum $k\in \Lambda _{N}^{\ast }$%
. The function $\varepsilon _{k}$ represents the kinetic energy and the real
number $\mu $ is the chemical potential.

The BCS interaction is defined via the BCS coupling function $\gamma _{k,q}$
which is usually assumed in the physics literature to be -- in momentum
space -- of the following form:%
\begin{equation}
\gamma _{k,q}=\left\{
\begin{array}{l}
\gamma \geq 0 \\
0%
\end{array}%
\begin{array}{l}
\mathrm{for\ }\left\Vert k-q\right\Vert \leq \mathrm{C} \\
\mathrm{for\ }\left\Vert k-q\right\Vert >\mathrm{C}%
\end{array}%
\right.   \label{physical choice}
\end{equation}%
with $\mathrm{C}\in \left( 0,\infty \right] $. The function $\gamma _{k,q}$
is not continuous in momentum space and so, it is slowly decaying, i.e.,
long range, in position space. (This means that it is not absolutely
summable.) The case $\varepsilon _{k}=0$ is known as the strong coupling
limit of the BCS model. Together with the choice $\mathrm{C}=\infty $, this
case is of interest, as its analysis is easier and at the same time it
qualitatively displays most of basic properties of real conventional type I
superconductors, see, e.g., Chapter VII, Section 4 in \cite{Thou}.

A general theory of superconductivity is, however, a subject of debate,
especially for high--$T_{c}$\emph{\ }superconductors. An important
phenomenon not taken into account in the BCS theory is the Coulomb
interaction between electrons or holes, which can imply strong correlations
-- for instance in high--$T_{c}$\emph{\ }superconductors. This problem was
of course already addressed in theoretical physics right after the emergence
of the Fr\"{o}hlich model and the BCS theory, see, e.g., \cite%
{Bogoliubov-tolman shirkov}. Most of theoretical methods are based on
perturbation theory or on the (non--rigorous) diagrammatic approach\ of
Quantum Field Theory. However, even if these approaches have been successful
in explaining many physical properties of superconductors \cite%
{Superconductivity2,Superconductivity3} only a few mathematically rigorous
results related to a microscopic description of the quantum many--body
problem exist as far as this problem is concerned, see for instance \cite%
{BruPedra1}.

Indeed, the results of \cite{BruPedra1} are based on an \emph{exact}
thermodynamic study of the phase diagram of the strong coupling BCS--Hubbard
model defined in a cubic box $\Lambda _{N}:=\{\mathbb{Z}\cap \lbrack
-L,L]\}^{D}$ of volume $|\Lambda _{N}|=N\geq 2$ by the Hamiltonian%
\begin{eqnarray}
\mathrm{H}_{N} &:&=-\mu \sum\limits_{x\in \Lambda _{N}}\left( n_{x,\uparrow
}+n_{x,\downarrow }\right) -h\sum_{x\in \Lambda _{N}}\left( n_{x,\uparrow
}-n_{x,\downarrow }\right)  \notag \\
&&+2\lambda \sum_{x\in \Lambda _{N}}n_{x,\uparrow }n_{x,\downarrow }-\frac{%
\gamma }{N}\sum_{x,y\in \Lambda _{N}}a_{x,\uparrow }^{\ast }a_{x,\downarrow
}^{\ast }a_{y,\downarrow }a_{y,\uparrow }  \notag \\
&&  \label{Hamiltonian BCS-Hubbard}
\end{eqnarray}%
for real parameters $\mu ,h\in \mathbb{R}$ and $\lambda ,\gamma \geq 0$. The
operator $a_{x,\mathrm{s}}^{\ast }$ (resp. $a_{x,\mathrm{s}}$) creates
(resp. annihilates) a fermion with spin $\mathrm{s}\in \{\uparrow
,\downarrow \}$ at lattice position $x\in \mathbb{Z}^{D}$, $D=1,2,3,...,$
whereas $n_{x,\mathrm{s}}:=a_{x,\mathrm{s}}^{\ast }a_{x,\mathrm{s}}$ is the
particle number operator at position $x$ and spin $\mathrm{s}$. The first
term of the right hand side of (\ref{Hamiltonian BCS-Hubbard}) represents
the strong coupling limit of the kinetic energy, also called
\textquotedblleft atomic limit\textquotedblright\ in the context of the
Hubbard model, see, e.g., \cite{atomiclimit1,atomiclimit2}. The second term
corresponds to the interaction between spins and the magnetic field $h$. The
one--site interaction with positive coupling constant $\lambda \geq 0$
represents the (screened) Coulomb repulsion as in the celebrated Hubbard
model. The last term is the BCS interaction written in the $x$--space since%
\begin{equation}
\frac{\gamma }{N}\sum_{x,y\in \Lambda _{N}}a_{x,\uparrow }^{\ast
}a_{x,\downarrow }^{\ast }a_{y,\downarrow }a_{y,\uparrow }=\frac{\gamma }{N}%
\sum_{k,q\in \Lambda _{N}^{\ast }}\tilde{a}_{k,\uparrow }^{\ast }\tilde{a}%
_{-k,\downarrow }^{\ast }\tilde{a}_{q,\downarrow }\tilde{a}_{-q,\uparrow },
\label{BCS interaction}
\end{equation}%
see (\ref{physical choice}) with $\mathrm{C}=\infty $. This homogeneous BCS
interaction should be seen as a long range effective interaction for which
the\ corresponding mediator does not matter, i.e., it could be due to
phonons, as in conventional type I superconductors, or anything else. From
theoretical considerations remark that phonons can probably not be
responsible of superconductivity above 90K, see, e.g., \cite{Emery}.

The Hamiltonian (\ref{Hamiltonian BCS-Hubbard}) defines in the thermodynamic
limit $N\rightarrow \infty $ a free energy density functional on a suitable
set of states of the fermionic observable algebra of the lattice $\mathbb{Z}%
^{D}$ ($D$--dimensional crystal). See \cite[Section 6.2]{BruPedra1} for
details. Minimizers $\omega $ of the free energy density are called
equilibrium states of the model. We say that the model has a superconducting
phase -- at fixed parameters -- if there is at least one equilibrium state $%
\omega $ of the model for which $\omega (a_{x,\downarrow }a_{x,\uparrow
})\neq 0$, i.e., if the $U(1)$--gauge symmetry is spontaneously broken.

The main properties of the phase diagram of the strong coupling BCS--Hubbard
model described in \cite{BruPedra1} can be summarized as follows:

\begin{itemize}
\item There is a non--empty set of parameters $\mathcal{S}$ defining a \emph{%
s--wave} superconducting phase with off--diagonal long range order.

\item Depending on parameters the superconducting phase transition is either
of first order or of second order (see Fig. \ref%
{domain-temp-critique-lamb.eps}).

\item The superconducting phase $\mathcal{S}$ is characterized by the
formation of Cooper pairs and a depleted Cooper pair condensate, the density
$\mathit{r}_{\beta }\in \lbrack 0,1/4]$ of which is defined by the gap
equation.

\item There is a Mei{\ss }ner effect concerning the relationship between
superconductivity and magnetization. The Mei{\ss }ner effect is defined here
by the absence of magnetization in presence of superconductivity. Steady
surface currents around the bulk of the superconductor are not analyzed as
it is a finite volume effect.

\item There exists a superconductor--Mott insulator phase transition for
near--one fermion densities per lattice site (see Fig. \ref%
{order-parameter-densite-1-1bis.eps}).

\item The coexistence of ferromagnetic and superconducting phases is shown
to be feasible at (critical) points of the boundary $\partial \mathcal{S}$
of $\mathcal{S}$.

\item The critical temperature $\theta _{c}$ can be (locally) an \emph{%
increasing} function of the \emph{positive} coupling constant $\lambda \geq
0 $ at fixed chemical potential $\mu \in \mathbb{R}$ (see Fig. \ref%
{domain-temp-critique-lamb.eps}), but not at fixed fermion density $\rho >0$.

\item For $\lambda \sim \gamma $ the critical temperature $\theta _{c}$
shows -- as a function of the fermion density $\rho $ -- the typical
\textquotedblleft superconducting domes\textquotedblright\ observed in high--%
$T_{c}$ superconductors: $\theta _{c}$ is zero or very small for $\rho \sim
1 $ and is much larger for $\rho $ away from $1$ (see Fig. \ref%
{order-parameter-densite-1-1bis.eps}). The latter gives a rigorous
explanation of the need of doping Mott insulators to obtain
superconductors.\
\end{itemize}

%TCIMACRO{%
%\TeXButton{figure domain-temp-critique-lamb.eps}{\begin{figure}[hbtp]
%\includegraphics[angle=0,scale=1,clip=true,width=6.5cm]{domain-temp-critique-lamb3bis}
%\caption{\emph{Illustration of the critical temperature $\theta _{c}$
%for $\gamma =2.6,$ $h=0$, $\mu =1.25$, and $\lambda \in [-0.1,0.85]$.
%The blue line corresponds to a second order phase transition,
%whereas the red dashed line represents the domain of $\lambda$ with a first order phase transition.}}
%\label{domain-temp-critique-lamb.eps}
%\end{figure}}}%
%BeginExpansion
\begin{figure}[hbtp]
\includegraphics[angle=0,scale=1,clip=true,width=6.5cm]{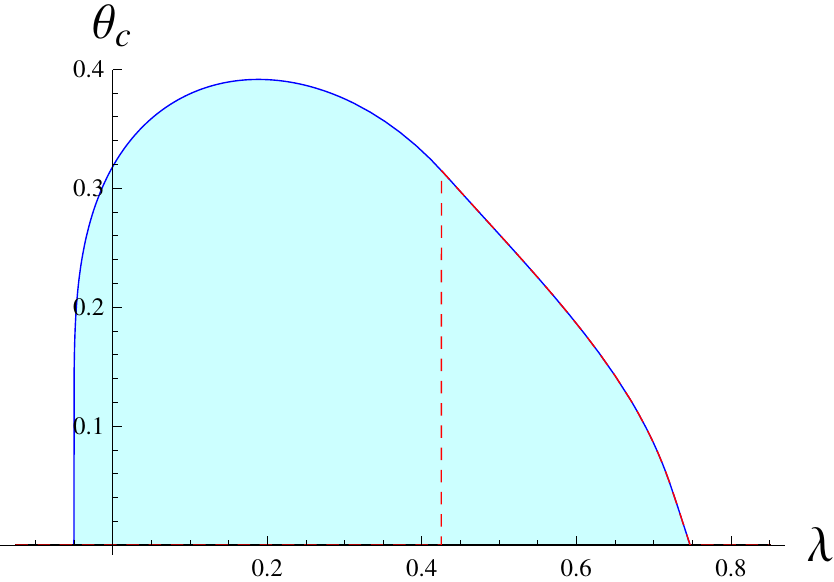}
\caption{\emph{Illustration of the critical temperature $\theta _{c}$
for $\gamma =2.6,$ $h=0$, $\mu =1.25$, and $\lambda \in [-0.1,0.85]$.
The blue line corresponds to a second order phase transition,
whereas the red dashed line represents the domain of $\lambda$ with a first order phase transition.}}
\label{domain-temp-critique-lamb.eps}
\end{figure}%
%EndExpansion
%TCIMACRO{%
%\TeXButton{figure order-parameter-densite-1-1bis.eps}{\begin{figure}[hbtp]
%\includegraphics[angle=0,scale=1,clip=true,width=6.5cm]{order-parameter-densite-1bisbis}
%\caption{\emph{Illustration of the critical temperature $\theta _{c}$
%for $\gamma =2.6,$ $h=0.1$, $\lambda =0.575$, and fermion densities $\rho \in [0,2]$.
%The blue and yellow region correspond respectively to the superconducting and ferromagnetic--superconducting phases.}}
%\label{order-parameter-densite-1-1bis.eps}
%\end{figure}}}%
%BeginExpansion
\begin{figure}[hbtp]
\includegraphics[angle=0,scale=1,clip=true,width=6.5cm]{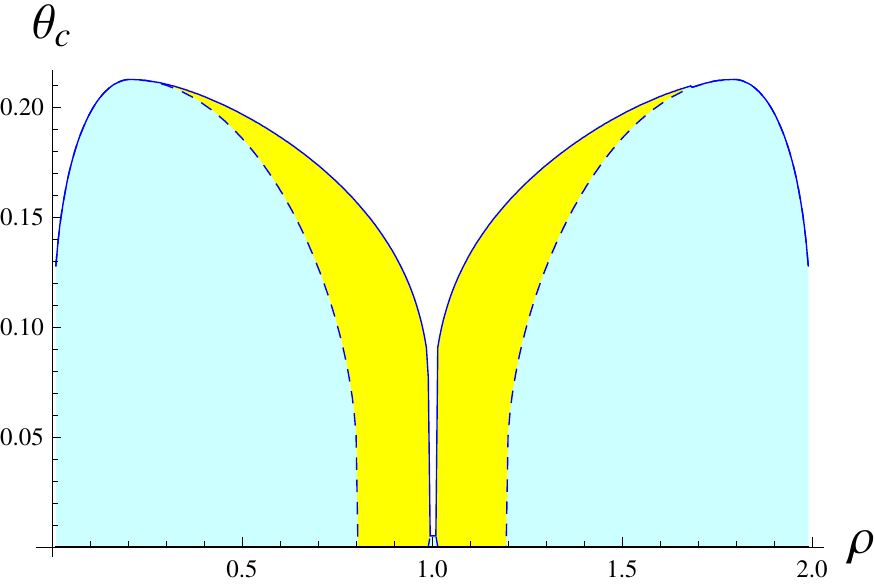}
\caption{\emph{Illustration of the critical temperature $\theta _{c}$
for $\gamma =2.6,$ $h=0.1$, $\lambda =0.575$, and fermion densities $\rho \in [0,2]$.
The blue and yellow region correspond respectively to the superconducting and ferromagnetic--superconducting phases.}}
\label{order-parameter-densite-1-1bis.eps}
\end{figure}%
%EndExpansion

This model is of course too simplified with respect to (w.r.t.) real
superconductors. For instance, the anti--ferromagnetic phase or the presence
of vortices, which can appear in (type II) high--$T_{c}$\emph{\ }%
superconductors, are not observed. However, since the range of parameters in
which we are interested turns out to be related to a first order phase
transition, by high--low temperature expansions the -- more realistic --
model including kinetic terms
\begin{equation}
\mathrm{H}_{N,\varepsilon }:=\mathrm{H}_{N}+\sum\limits_{x,y\in \Lambda
_{N}}\varepsilon (x-y)\left( a_{y,\downarrow }^{\ast }a_{x,\downarrow
}+a_{y,\uparrow }^{\ast }a_{x,\uparrow }\right)
\label{model with kinetic energy}
\end{equation}%
should have essentially the same correlation functions as $\mathrm{H}_{N}$
-- up to corrections of order $||\varepsilon ||_{1}$ ($\ell ^{1}$--norm of $%
\varepsilon $). Thus, the Hamiltonian $\mathrm{H}_{N}$ may be a good model
for certain kinds of superconductors or ultra--cold Fermi gases in optical
lattices for which the strong coupling regime is justified, especially
because all parameters of $\mathrm{H}_{N}$ have a phenomenological
interpretation and can be directly related to experiments, see \cite[Sect. 5]%
{BruPedra1}.

Since the discovery of mercury superconductivity in 1911 a significant
amount of superconducting materials has been found. This includes usual
metals, like lead, aluminum, zinc or platinum, magnetic materials,
heavy--fermion systems, organic compounds and ceramics. High--$T_{c}$\emph{\
}cuprate\emph{\ }superconductors are among the most interesting
superconducting materials for applications. In spite of that, a general
microscopic theory explaining conveniently their thermodynamics is still not
available. For instance, the experimentally well--known electron--hole
asymmetry of high--$T_{c}$\emph{\ }superconductors which usually show higher
critical temperatures for donors of holes than for donors of electrons, is
not clearly understood from the microscopic point of view. As far as we know
no microscopic theory can give up to now some incontestable explanation of
this fact unless one artificially breaks the electron--hole symmetry of
models (for instance by imposing explicitly two different mass for electrons
and holes). In fact, the most plausible explanation of the electron--hole
asymmetry is given in \cite{Superconductivity2,Superconductivity3} and is
also based on two--band (Hubbard or $t-J$) models in the strong coupling
regime. These studies are debatable and we discuss them in relation with our
approach in Section \ref{section II} after introducing our two--band model.

The usual BCS model and the strong coupling BCS--Hubbard model have
structural properties preventing their critical temperatures of
superconductivity from being asymmetric w.r.t. the density of donors. The
strong coupling BCS--Hubbard shows -- at least and in contrast to the usual
BCS model -- the typical (though symmetric) \textquotedblleft
superconducting domes\textquotedblright\ of high--$T_{c}$ materials for the
critical temperature. Therefore, the first objective is to investigate
generalizations of the strong coupling BCS--Hubbard in order to find
plausible -- microscopic -- explanations for the electron--hole asymmetry in
high--$T_{c}$\emph{\ }cuprate superconductors. The (one--band) strong
coupling BCS--Hubbard is a good starting point for further investigations on
high--$T_{c}$ phenomenology, also because it correctly describes the
non--superconducting phase of cuprates near half--filling, which is
Mott--insulating and not metallic as in usual superconductors.

Another class of superconductors which would be interesting to study are
ferromagnetic superconductors \cite[p. 263-267]{Shrivastava}. Indeed, these
materials exhibit some peculiar features not found in ordinary
superconductors, one of these being the reentering behavior: The system
becomes superconducting below a critical temperature $\theta _{c_{1}}$ and
then magnetically ordered but not superconducting below $\theta _{c_{2}}$
with $\theta _{c_{2}}<\theta _{c_{1}}$. Additionally, coexistence of a
ferromagnetic phase and superconductivity seems also to appear around $%
\theta _{c_{2}}$, at least for some ferromagnetic superconductors. Motivated
by the thermodynamic behavior of ferromagnetic superconductors, the second
goal of this paper is to show that a magnetic inter--band interaction can be
responsible for a reentering behavior.

We introduce a two--band model by using two copies of the (one--band) model $%
\mathrm{H}_{N}$ defined by (\ref{Hamiltonian BCS-Hubbard}) and by adding an
inter--band interaction term. The parameters of the first band, called here
the \textquotedblleft $\mathit{s}$\textquotedblright --band, are chosen such
that a superconducting phase appears at low enough temperatures if no
interaction with the second band is present. By contrast, the parameters of
the second band, called the \textquotedblleft $\mathit{f}$\textquotedblright
--band, are such that without interaction with the \textquotedblleft $%
\mathit{s}$\textquotedblright --band one would have a (generally
ferromagnetic) Mott--insulating phase at low temperatures.

We prove that the thermodynamic behavior of such a two--band model inherits
of course properties of the thermodynamic behavior of the one--band model $%
\mathrm{H}_{N}$ described in \cite{BruPedra1}, but also shows additional
features:

\begin{itemize}
\item There is an electron--hole asymmetry w.r.t. the critical temperature
of superconductivity which results from the inter--band interaction and
intra--band Coulomb repulsion in interplay with thermal fluctuations between
the \textquotedblleft $\mathit{s}$\textquotedblright -- and
\textquotedblleft $\mathit{f}$\textquotedblright --bands.

\item Choosing the relative strength of parameters to each other according
to experimental data about the energy levels of holes in $\mathrm{CuO}_{2}$
layers of high--$T_{c}$\emph{\ }cuprate\emph{\ }superconductors,\ we show
that this asymmetry is in favor of a doping with donors of holes. This is
just what is seen in cuprates.

\item There is a (rather small) set of parameters with a reentering behavior
coming from a magnetic inter--band interaction.

\item There is a kind of \textquotedblleft microscopic Mei{\ss }ner
effect\textquotedblright : The increase of the magnetic inter--band
interaction destroys superconductivity without need of any magnetization
induced by an external magnetic field. It follows that a reentering behavior
can also appear without any ferromagnetism.
\end{itemize}

Our two--band model is certainly not the final microscopic theory of high--$%
T_{c}$\emph{\ }cuprate\emph{\ }superconductors even if one includes a small
kinetic part as explained above. However, we are convinced that the model
and analysis presented here highlight microscopic processes playing a
crucial role in which concerns the phenomenology of high--$T_{c}$ materials
-- or at least a relevant part of it.

The method presented in \cite{BruPedra1} and used here gives access to
domains of the phase diagram usually difficult to reach via other standard
mathematical tools. For instance, the existence of the superconductor--Mott
insulator phase transition for near--one fermion densities per lattice site
(see Fig. \ref{order-parameter-densite-1-1bis.eps}) can neither be obtained
by perturbation theory nor by spin reflection positivity \cite%
{Lieb-reflexion} arguments. Indeed, the spin reflection positivity described
for instance in \cite{Tian} cannot give access to such a phenomenon as it
requires a fermion density exactly equal to one. Also, the regime in which
the superconducting domes are observed corresponds to the choice $\lambda
\sim \gamma $ and they are never seen if $\lambda $ is too small as compared
to $\gamma $. Thus, perturbative arguments around the purely BCS case $%
\lambda =0$ are inadequate. The results given here may be rigorously
obtained by renormalization group techniques, but this method of analysis
would be -- from the technical point of view -- probably much more demanding
than the approach of \cite{BruPedra1}.

As a final remark the one--band and two--band models discussed here should
be seen as a kind of \textquotedblleft solvable\textquotedblright\ class of
models capturing important phenomenological aspects and from which one can
implement physically more realistic models -- for instance by using
perturbation theory. Indeed, a first natural extension of the two--band
model -- beyond the introduction of a small kinetic term as explained above
-- is the addition of a Heisenberg--type interaction for fermions within the
non--superconducting band in order to spontaneously create a ferromagnetic
phase and to get a reasonable microscopic theory of ferromagnetic
superconductors. Another possible generalization is the introduction of a
small inter--band interaction term describing tunneling effects between
bands (inter--band hopping), see below (\ref{tunnel term}).

\section{The two--band model\label{section II}}

In order to fix ideas about the microscopic structure of typical high--$T_{c}
$ materials we consider the case of cuprates. It is known from experimental
physics that superconducting carriers, mainly holes in the case of cuprates,
move within two--dimensional $\mathrm{CuO}_{2}$ layers made of $\mathrm{Cu}%
^{++}$ and $\mathrm{O}^{--}$, see, e.g., \cite[Fig. 5.3. p. 127]{Saxena}.

$\mathrm{Cu}^{+}=(\mathrm{Ar})3d^{10}$ is a closed shell configuration (see,
e.g., \cite{Emery} or \cite[Sect. 5.5]{Saxena}). Then the last occupied
orbital of the copper cations $\mathrm{Cu}^{++}=(\mathrm{Ar})3d^{9}$ is
almost full, i.e., it can be modelled by a \emph{half--filled}\ band of
holes (exactly one hole per lattice site). Putting a further hole in one
copper cation, i.e., transforming $\mathrm{Cu}^{++}$ into $\mathrm{Cu}^{+++}$%
, costs some positive amount $2\lambda _{\mathrm{Cu}}>0$ of energy. The
latter justifies the presence of a repulsion term of the form
\begin{equation*}
2\lambda _{\mathrm{Cu}}n_{x,\uparrow }^{(\mathrm{Cu})}n_{x,\downarrow }^{(%
\mathrm{Cu})}\quad \mathrm{with\ }\lambda _{\mathrm{Cu}}>0
\end{equation*}%
at each (two--dimensional) lattice site $x$ for the copper band.

$\mathrm{O}^{--}=(\mathrm{Be})2p^{6}$ corresponds to a closed shell
configuration as well (see, e.g., \cite{Emery} or \cite[Sect. 5.5]{Saxena})
and therefore, the oxygen orbitals in the $\mathrm{CuO}_{2}$ layers of
cuprates can be modelled by an \emph{empty}\ band of holes provided there is
no doping. Indeed, by doping cuprates holes can be added to orbitals of $%
\mathrm{O}^{--}$ in the $\mathrm{CuO}_{2}$ layers. Putting one new hole on
an oxygen site analogously costs some positive amount $\delta _{\mathrm{Cu-O}%
}>0$ of energy relatively to the energy of the half--filled $\mathrm{Cu}%
^{++} $ orbital. This justifies the assumption that the difference between
chemical potentials $\mu _{\mathrm{Cu}}$ and $\mu _{\mathrm{O}}$ of the
copper and oxygen bands is some strictly positive quantity $\delta _{\mathrm{%
Cu-O}}>0$, i.e.,
\begin{equation}
\mu _{\mathrm{Cu}}=\mu _{\mathrm{O}}+\delta _{\mathrm{Cu-O}}>\mu _{\mathrm{O}%
}\ .  \label{difference chemical potential0}
\end{equation}%
Experiments on cuprates indicate that $\lambda _{\mathrm{Cu}}\approx 4\rm{%
eV}$ and $\delta _{\mathrm{Cu-O}}\leq 2\rm{eV}$ (at least in \cite[Fig.
5.9. p. 132]{Saxena}). So, $\delta _{\mathrm{Cu-O}}$ should be chosen small
as compared to the intra--band repulsion $\lambda _{\mathrm{Cu}}$. At any
site $x$ of the $\mathrm{CuO}_{2}$ lattice it is also natural to assume that
two carriers experience an intra--band repulsion
\begin{equation*}
2\lambda _{\mathrm{O}}n_{x,\uparrow }^{(\mathrm{O})}n_{x,\downarrow }^{(%
\mathrm{O})}\quad \mathrm{with\ }\lambda _{\mathrm{O}}\geq 0
\end{equation*}%
when they are both on the oxygen \ site and an inter--band repulsion
\begin{equation*}
2\lambda _{\mathrm{Cu-O}}(n_{x,\uparrow }^{(\mathrm{O})}+n_{x,\downarrow }^{(%
\mathrm{O})})(n_{x,\uparrow }^{(\mathrm{Cu})}+n_{x,\downarrow }^{(\mathrm{Cu}%
)})\quad \mathrm{with\ }\lambda _{\mathrm{Cu-O}}\geq 0
\end{equation*}%
if one carrier is on the oxygen site and the other one on the copper site,
see, e.g., \cite[Eq. (1)]{Emery}. More precise experimental data about the
strength of the couplings $\lambda _{\mathrm{O}}$ and $\lambda _{\mathrm{Cu-O%
}}$ in real cuprates would be useful.

To each of these (almost independent) $\mathrm{CuO}_{2}$ conducting layers
there is a second layer of another chemical nature which acts as a charge
reservoir and provides conduction fermions to the $\mathrm{CuO}_{2}$ layer.
These reservoir layers can be engineered in order to get the desired density
of carriers (holes or electrons). In $\mathrm{La}_{2-x}\mathrm{Sr}_{x}%
\mathrm{CuO}_{4}$ compounds the holes added to $\mathrm{CuO}_{2}$ layers of $%
\mathrm{La}_{2}\mathrm{CuO}_{4}$ directly go into the oxygen band which is
experimentally shown to be at the origin of superconductivity. Indeed, holes
on copper bands cannot directly participate in superconductivity by forming
Cooper pairs, see \cite[p. 133]{Saxena}. Additionally, the hopping amplitude
of holes within $\mathrm{CuO}_{2}$ layers of $\mathrm{La}_{2-x}\mathrm{Sr}%
_{x}\mathrm{CuO}_{4}$ is smaller or equal than $1\rm{eV}\leq \delta _{%
\mathrm{Cu-O}}\leq 2\lambda _{\mathrm{Cu}}$. In particular, the assumption
of strong coupling regime is far from being unrealistic in this case. For
more details we recommend \cite{Saxena}, in particular Chapter 5.

As suggested in \cite[Sect. 5.12]{Saxena} we define a two--band model in
relation to the physics described above. We use two lattices $\mathfrak{L}_{%
\mathit{s}}$ and $\mathfrak{L}_{\mathit{f}}$ corresponding respectively to a
\textquotedblleft $\mathit{s}$\textquotedblright --band\ of superconducting
fermions and a \textquotedblleft $\mathit{f}$\textquotedblright --band of\
ferromagnetic fermions (electrons, holes or even more general fermions, like
fermionic magnetic atoms). In particular, the \textquotedblleft $\mathit{s}$%
\textquotedblright --band is interpreted within $\mathrm{CuO}_{2}$
conducting layers in cuprates as the oxygen band, whereas the
\textquotedblleft $\mathit{f}$\textquotedblright --band corresponds to the
copper band. To simplify our study, we assume that both lattices are of the
same type: $\mathfrak{L}_{\mathit{s}}\sim \mathfrak{L}_{\mathit{f}}\sim
\mathbb{Z}^{D}$, $D=1,2,3,...$ In this context the thermodynamics of the
two--band system is governed by the (strong coupling) Hamiltonian%
\begin{equation}
\mathrm{H}_{N}^{(\mathit{s},\mathit{f})}:=\mathrm{H}_{N}^{(\mathit{s}%
)}\otimes \mathbf{1}+\mathbf{1}\otimes \mathrm{H}_{N}^{(\mathit{f})}+\mathrm{%
V}_{N}  \label{fullhamilton}
\end{equation}%
where $N=|\Lambda _{N}|<\infty $ is the volume of the box $\Lambda _{N}:=\{%
\mathbb{Z}\cap \lbrack -L,L]\}^{D}$ seen as a subset of either $\mathfrak{L}%
_{\mathit{s}}$ or $\mathfrak{L}_{\mathit{f}}$. Here $\mathbf{1}$ is the
identity on the fermion Fock space related to either $\mathfrak{L}_{\mathit{s%
}}$ or $\mathfrak{L}_{\mathit{f}}$. In this model the kinetic energy is
neglected as the strong coupling regime is realistic for $\mathrm{CuO}_{2}$
layers. In fact, without changing qualitatively the phenomenology a small
kinetic energy could be added as explained around (\ref{model with kinetic
energy}). Accordingly, the self--adjoint operator
\begin{equation*}
\mathrm{H}_{N}^{(\mathit{s})}:=\mathrm{H}_{N}\left( \mu _{\mathit{s}},h_{%
\mathit{s}},\lambda _{\mathit{s}},\gamma _{\mathit{s}}\right)
\end{equation*}%
is the Hamiltonian of \textquotedblleft $\mathit{s}$\textquotedblright
--fermions defined from $\mathrm{H}_{N}$ with $\mu _{\mathit{s}},h_{\mathrm{s%
}}\in \mathbb{R}$ and $\lambda _{\mathit{s}},\gamma _{\mathit{s}}\geq 0$,
whereas
\begin{equation*}
\mathrm{H}_{N}^{(\mathit{f})}:=\mathrm{H}_{N}\left( \mu _{\mathit{f}},h_{%
\mathit{f}},\lambda _{\mathit{f}},0\right)
\end{equation*}%
is the Hamiltonian of \textquotedblleft $\mathit{f}$\textquotedblright
--fermions with $\mu _{\mathit{f}},h_{\mathit{f}}\in \mathbb{R}$ and $%
\lambda _{\mathit{f}}\geq 0$. Similarly to the discussion above about the
energy levels in copper and oxygen bands of $\mathrm{CuO}_{2}$ layers, we
write
\begin{equation}
\mu _{\mathit{f}}:=\mu _{\mathit{s}}+\delta \quad \mathrm{with}\ \delta \in
\mathbb{R}.  \label{difference chemical potential}
\end{equation}%
In particular, if $\delta >0$, then the \textquotedblleft $\mathit{f}$%
\textquotedblright --band\ is energetically lower than the \textquotedblleft
$\mathit{s}$\textquotedblright --band,\ whereas $\delta <0$ means the
opposite. As far as $\mathrm{CuO}_{2}$ layers are concerned the energy
difference $\delta =\delta _{\mathrm{Cu-O}}>0$ must be positive and small
w.r.t. the repulsive coupling constant $\lambda _{\mathit{f}}=\lambda _{%
\mathrm{Cu}}\geq 0$. The inter--band interaction equals
\begin{eqnarray}
\mathrm{V}_{N} &:&=2\lambda \sum_{x\in \Lambda _{N}}(n_{x,\uparrow }^{(%
\mathit{s})}+n_{x,\downarrow }^{(\mathit{s})})(n_{x,\uparrow }^{(\mathit{f}%
)}+n_{x,\downarrow }^{(\mathit{f})})  \notag \\
&&-\eta \sum_{x\in \Lambda _{N}}(n_{x,\uparrow }^{(\mathit{s}%
)}-n_{x,\downarrow }^{(\mathit{s})})(n_{x,\uparrow }^{(\mathit{f}%
)}-n_{x,\downarrow }^{(\mathrm{f})})  \label{hamiltoninteraction}
\end{eqnarray}%
with $\lambda \geq 0$ and $\eta \in \mathbb{R}$. Here $n_{x,\mathrm{s}}^{(%
\mathit{s})}$ (resp. $n_{x,\mathrm{s}}^{(\mathit{f})}$) is the particle
number operator of \textquotedblleft $\mathit{s}$\textquotedblright --
(resp. \textquotedblleft $\mathit{f}$\textquotedblright --) fermions at
position $x\in \mathbb{Z}^{D}$ and spin $\mathrm{s}\in \{\uparrow
,\downarrow \}$. The first term of the inter--band interaction $\mathrm{V}%
_{N}$ represents the screened Coulomb interaction between \textquotedblleft $%
\mathit{s}$\textquotedblright --\ and \textquotedblleft $\mathit{f}$%
\textquotedblright --fermions, i.e., between carriers in the copper and
oxygen bands of the $\mathrm{CuO}_{2}$ conducting layers. If $\eta \neq 0$
then the second term in $\mathrm{V}_{N}$ represents a magnetic interaction
between \textquotedblleft $\mathit{s}$\textquotedblright -- and
\textquotedblleft $\mathit{f}$\textquotedblright --fermions on the same site
of the lattices $\mathfrak{L}_{\mathit{s}}$ and $\mathfrak{L}_{\mathit{f}}$.
\

Note that the superconducting phase which appears for \textquotedblleft $%
\mathit{s}$\textquotedblright --fermions in our two--band model, is a
(perfect) \emph{s--wave} phase (cf. \cite[Thm 3.3]{BruPedra1}) for all $%
D=1,2,3,...$ The dimension $D=2$ of conducting layers is of course extremely
important for the physics of high--$T_{c}$\emph{\ }cuprate\emph{\ }%
superconductors. Moreover, experiments usually suggest that \emph{d--wave}
superconductivity can appear in cuprate high--$T_{c}$ superconductors and
this is an important physical aspect of high--$T_{c}$ superconductors. The
latter as well as the anti-ferromagnetism in cuprates are not studied here.
However, as the phenomenology which emerges is coherent with experiments,
the dimensionality of conducting layers, the anti-ferromagnetism, and the
kind of pairing of superconducting carriers do not seem to be relevant for
the phenomenon we are interested in, namely the influence of multi--band
structures on the phase diagram of generic superconductors.

For instance, the phenomenon we present here highlight why the explanation
given in \cite{Superconductivity2,Superconductivity3} of the electron-hole
asymmetry works. This approach is, indeed, based on a two-band model in the
strong coupling regime. However, in contrast to our approach, the two bands
in \cite{Superconductivity2,Superconductivity3} come from two different
hopping terms, representing respectively the nearest--neighbor and
next--nearest--neighbor hopping. Their studies are far from being
mathematically rigorous and can thus be debatable. Our paper is a
mathematical proof that the kinetic energy is not necessary to explain this
asymmetry and shows the importance of close multi-band structures (even in
presence of small kinetic terms as explained around (\ref{model with kinetic
energy}) and (\ref{fullhamilton})). In fact, even if \cite%
{Superconductivity2,Superconductivity3} is physically correct, the
explanations \cite{Superconductivity2,Superconductivity3} are not
transparent (at least to the non--expert) and we think that our paper gives
a much clearer understanding why that should work, where this asymmetry
comes from, and how that can be used to draw different conclusions (cf.
Section \ref{Section III}, cases (a)--(d)).

Observe also that the bands are denoted by \textquotedblleft $\mathit{s}$%
\textquotedblright\ for \textquotedblleft superconducting\textquotedblright\
and by \textquotedblleft $\mathit{f}$\textquotedblright\ for
\textquotedblleft ferromagnetic\textquotedblright\ or \textquotedblleft
fixed\textquotedblright . The charge carriers can be here fermions of any
kind. In cuprates such fermions should be seen as holes (of electrons).
Indeed, the two--band model can emerge from other physical systems or
interpretations. For instance, the \textquotedblleft $\mathit{s}$%
\textquotedblright --band could have represented holes within $\mathrm{CuO}%
_{2}$ layers, whereas \textquotedblleft $\mathit{f}$\textquotedblright
--fermions\ could have been seen as holes within the charge reservoir layer.
Another example could have been given by considering some tunneling between
the conducting layer and the charge reservoir layer. The tunneling modes
could then give origin to different effective energy levels. The latter can
be interpreted as a discrete component of the kinetic energy in the
direction perpendicular to the conduction plane. Indeed, we can assume that
the layers of high--$T_{c}$ materials can be decomposed in independent
groups of finitely many layers (for instance groups of two layers in the
case of cuprates: One $\mathrm{CuO}_{2}$ layer and one reservoir layer).
Using periodic boundary conditions the kinetic energy w.r.t. the
perpendicular direction to the conduction planes is -- in this case -- a
discrete quantity and leads to the formation of finitely many different
energetic levels per site. With this example the \textquotedblleft $\mathit{s%
}$\textquotedblright --band \ would have represented holes or electrons with
a zero perpendicular kinetic energy, whereas the \textquotedblleft $\mathit{f%
}$\textquotedblright --band\ would have corresponded to holes or electrons
moving with a velocity having a non--vanishing component perpendicularly to
the conduction planes.

Now, the first question we have to address in order to obtain the
thermodynamics of the system is the computation of the grand--canonical
pressure%
\begin{equation*}
\mathrm{p}:=\underset{N\rightarrow \infty }{\lim }\frac{1}{\beta N}\ln
\mathrm{Trace}\left( e^{-\beta \mathrm{H}_{N}^{(\mathit{s},\mathit{f}%
)}}\right) <\infty
\end{equation*}%
in the thermodynamic limit $N\rightarrow \infty $ at fixed inverse
temperature $\beta >0$. Such an analysis is performed rigorously in \cite%
{BruPedra1} for the one--band case (see also Section \ref{appendix}) and can
easily be extended to the two--band model $\mathrm{H}_{N}^{(\mathit{s},%
\mathit{f})}$. Using the methods of \cite{BruPedra1} (see Section \ref%
{appendix}) together with explicit computations we get the following result:

\begin{theorem}
\label{BCS theorem 1}For any real $\mu _{\mathit{s}}$, $h_{\mathit{s}}$, $%
\mu _{\mathit{f}}$, $h_{\mathit{f}}$, $\eta $ and positive numbers $\gamma _{%
\mathit{s}}$, $\lambda _{\mathit{s}}$, $\lambda _{\mathit{f}}$, $\lambda $, $%
\beta >0$,
\begin{equation*}
\mathrm{p}:=\beta ^{-1}\ln 2+\mu _{\mathit{s}}+\mu _{\mathit{f}}+\sup_{r\geq
0}\left\{ -\gamma _{\mathit{s}}r+\beta ^{-1}\ln \mathrm{f}(r)\right\}
\end{equation*}%
with%
\begin{eqnarray*}
\mathrm{f}(r) &=&(\mathrm{e}^{-\beta \mu _{\mathit{f}}}+\mathrm{e}^{-\beta
(4\lambda +2\lambda _{\mathit{f}}-\mu _{\mathit{f}})})\cosh (\beta h_{%
\mathit{s}}) \\
&&+\mathrm{e}^{-\beta (2\lambda -h_{\mathit{s}})}\cosh (\beta (\eta +h_{%
\mathit{f}})) \\
&&+\mathrm{e}^{-\beta (2\lambda +h_{\mathit{s}})}\cosh (\beta (\eta -h_{%
\mathit{f}})) \\
&&+\mathrm{e}^{-\beta (\lambda _{\mathit{s}}+\mu _{\mathit{f}})}\cosh (\beta
g_{r,0}) \\
&&+\mathrm{e}^{-\beta (4\lambda +\lambda _{\mathit{s}}+2\lambda _{\mathit{f}%
}-\mu _{\mathit{f}})}\cosh (\beta g_{r,4\lambda }) \\
&&+2\mathrm{e}^{-\beta (2\lambda +\lambda _{\mathit{s}})}\cosh (\beta h_{%
\mathit{f}})\cosh (\beta g_{r,2\lambda })
\end{eqnarray*}%
and $g_{r,x}:=\{(x+\lambda _{\mathit{s}}-\mu _{\mathit{s}})^{2}+\gamma _{%
\mathit{s}}^{2}r\}^{1/2}$.
\end{theorem}

The exact form of the function $\mathrm{f}$ is only given for completeness.
What is important is that it is some explicitly given function in the
thermodynamic limit -- even if we do not know, a priori, how to diagonalize
the Hamiltonian $\mathrm{H}_{N}^{(\mathit{s},\mathit{f})}$ at any fixed $%
N\in \mathbb{N}$. This fact is important as it makes possible a computer
aided rigorous analysis of the thermodynamic problem. There is, however, a
certain amount on luckiness about this: The function $\mathrm{f}$ is
obtained from the eigenvalues of some $(16\times 16)$--matrix, the
so--called approximating Hamiltonian of the model (see \cite{BruPedra1} and
Section V for details), depending on the parameters of the model. By a
well--known theorem of algebra there is no explicit general formula for the
solutions of polynomial equations of degree greater than four. So, it is a
rather surprising property of the characteristic polynomial (which has
degree $16$) of the matrix corresponding to the approximating Hamiltonian of
our problem to have explicitly known zeros for any choice of parameters.

In fact, it would have been natural to include in our model $\mathrm{H}%
_{N}^{(\mathit{s},\mathit{f})}$ a hopping term of the form
\begin{equation}
\sum\limits_{x\in \Lambda _{N},\mathrm{s}\in \{\uparrow ,\downarrow \}}t_{%
\mathit{s},\mathit{f}}\left( (a_{x,\mathrm{s}}^{(\mathit{s})})^{\ast }a_{x,%
\mathrm{s}}^{(\mathit{f})}+(a_{x,\mathrm{s}}^{(\mathit{f})})^{\ast }a_{x,%
\mathrm{s}}^{(\mathit{s})}\right)  \label{tunnel term}
\end{equation}%
which on all lattice sites $x\in \Lambda _{N}$ annihilates a fermion in
\textquotedblleft $\mathit{f}$\textquotedblright --band to create another
one within the \textquotedblleft $\mathit{s}$\textquotedblright --band and
vice--versa. Theorem \ref{BCS theorem 1} would still be satisfied for any $%
t_{\mathit{s},\mathit{f}}\in \mathbb{R}$, but the function $\mathrm{f}$
would then be more difficult to obtain as the eigenvalues of the resulting $%
(16\times 16)$--matrix are not that easy to compute. However, the hopping
term (\ref{tunnel term}), which is important in the analysis dynamical
properties, should have almost no effect on the thermodynamics of the system
at equilibrium -- at least if $|t_{\mathit{s},\mathit{f}}|$ is sufficiently
small. This conjecture could anyway be verified -- but with a significant
numerical effort on the diagonalization of the corresponding $(16\times 16)$%
--matrix.

To conclude, let $\mathfrak{R}\subseteq \lbrack 0,\infty )$ be the set of
solutions $\mathit{r}_{\beta }\geq 0$ of the variational problem
\begin{equation}
\underset{r\geq 0}{\sup }\left\{ -\gamma _{\mathit{s}}r+\beta ^{-1}\ln
\mathrm{f}(r)\right\} =-\gamma _{\mathit{s}}\mathit{r}_{\beta }+\beta
^{-1}\ln \mathrm{f}(\mathit{r}_{\beta })  \label{BCS pressure 2}
\end{equation}%
given by Theorem \ref{BCS theorem 1}. Observe that $\mathfrak{R}\neq
\varnothing $ is a non--empty compact set since $\mathrm{f}(r)=\mathcal{O}(%
\mathrm{e}^{\beta \gamma _{\mathit{s}}\sqrt{r}})$ when $r\rightarrow \infty $%
. Let
\begin{equation*}
\mathfrak{E}:=\{z_{\beta }=\sqrt{\mathit{r}_{\beta }}e^{i\varphi }:\varphi
\in \lbrack 0,2\pi ),\;\mathit{r}_{\beta }\in \mathfrak{R}\}\subset
%TCIMACRO{\U{2102} }%
%BeginExpansion
\mathbb{C}
%EndExpansion
.
\end{equation*}%
Exactly as in the case of the one--band model $\mathrm{H}_{N}$ (cf. \cite[%
Section 6.2]{BruPedra1}) the set $\mathfrak{E}$ parameterizes (one--to--one)
the pure equilibrium states of the model $\mathrm{H}_{N}^{(\mathit{s},%
\mathit{f})}$. The equilibrium state $\omega $ is called pure, if $\omega
=\lambda \omega _{1}+(1-\lambda )\omega _{2}$ for equilibrium states $\omega
_{1},\omega _{2}$ and some $\lambda \in (0,1)$ implies $\omega _{1}=\omega
_{2}=\omega $. The parameterization $\mathfrak{E}\ni z_{\beta }\mapsto
\omega _{z_{\beta }}$ can be chosen to be continuous and such that $\omega
_{z_{\beta }}(a_{x,\downarrow }^{(\mathit{s})}a_{x,\uparrow }^{(\mathit{s}%
)})=z_{\beta }$ for all $x\in
%TCIMACRO{\U{2124} }%
%BeginExpansion
\mathbb{Z}
%EndExpansion
^{D}$. In particular, if $\mathit{r}_{\beta }\in $ $\mathfrak{R}$, $\mathit{r%
}_{\beta }>0$, then the model $\mathrm{H}_{N}^{(\mathit{s},\mathit{f})}$ has
a superconducting phase for the corresponding set of parameters.

As in the one--band case any (weak$^{\ast }$--) limit point of the local
Gibbs states%
\begin{equation}
\omega _{N}\left( \cdot \right) :=\frac{\mathrm{Trace}\left( \mathrm{\ }%
\cdot \mathrm{\ }e^{-\beta \mathrm{H}_{N}^{(\mathit{s},\mathit{f})}}\right)
}{\mathrm{Trace}\left( e^{-\beta \mathrm{H}_{N}^{(\mathit{s},\mathit{f}%
)}}\right) }  \label{BCS gibbs state Hn}
\end{equation}%
associated with $\mathrm{H}_{N}^{(\mathit{s},\mathit{f})}$ is an equilibrium
state (cf. \cite[Thm 6.5]{BruPedra1}). As all pure equilibrium states $%
\{\omega _{z_{\beta }}\}_{z_{\beta }\in \mathfrak{E}}$ can be explicitly
described and any equilibrium state is some convex combination of the pure
ones, we obtain a direct description, at once, of all correlation functions
of the Gibbs state $\omega _{N}$ in the limit $N\rightarrow \infty $. We do
not enter into the detailed proofs of these facts here because they are
essentially the same as for the one--band case (cf. \cite[Section 6]%
{BruPedra1}). The only important thing to keep in mind is that the local
Gibbs state $\omega _{N}$ is explicitly known in the limit $N\rightarrow
\infty $. Hence we have access to the entire phase diagram of the two--band
model which is the main achievement of the present paper.

\section{Existence of asymmetric superconducting domes\label{Section III}}

The first important thing we would like to analyze here is the influence of
a \textquotedblleft $\mathit{f}$\textquotedblright --band on
superconductivity of the \textquotedblleft $\mathit{s}$\textquotedblright
--band via the screened Coulomb interaction
\begin{equation}
2\lambda \sum_{x\in \Lambda _{N}}(n_{x,\uparrow }^{(\mathit{s}%
)}+n_{x,\downarrow }^{(\mathit{s})})(n_{x,\uparrow }^{(\mathit{f}%
)}+n_{x,\downarrow }^{(\mathit{f})})  \label{coulomb repulsionbis}
\end{equation}%
between fermions of both bands.

As explained above it is natural to relate the chemical potentials $\mu _{%
\mathit{s}}$ and $\mu _{\mathit{f}}$ by the energy difference between bands,
i.e., $\mu _{\mathit{f}}:=\mu _{\mathit{s}}+\delta $ with $\delta \in
\mathbb{R}$, see (\ref{difference chemical potential0}) and (\ref{difference
chemical potential}). If $\left\vert \delta \right\vert $ is sufficiently
large then it is expected that the \textquotedblleft $\mathit{f}$%
\textquotedblright --band\ has no effect on superconductivity of
\textquotedblleft $\mathit{s}$\textquotedblright --fermions. As a
consequence we concentrate our study on two--band systems with small energy
gaps $\left\vert \delta \right\vert $. We divide our analysis at fixed
chemical potential $\mu _{\mathit{s}}$ and energy gap $\delta $ in four main
cases:

\begin{itemize}
\item[(a)] Effect of an almost empty \textquotedblleft $\mathit{f}$%
\textquotedblright --band\ ($\delta <0$) on superconductivity.

\item[(b)] Breakdown of the half--filled \textquotedblleft $\mathit{f}$%
\textquotedblright --band destroying superconductivity.

\item[(c)] Breakdown of the half--filled \textquotedblleft $\mathit{f}$%
\textquotedblright --band implying superconductivity. This should be the
case of $\mathrm{CuO}_{2}$ layers in cuprates.

\item[(d)] Effect of an\ almost full\ \textquotedblleft $\mathit{f}$%
\textquotedblright --band\ ($\delta >0$) on superconductivity.
\end{itemize}

\noindent Before starting this program it is necessary to precise different
thermodynamic functions. First, we recall that the solution $\mathit{r}%
_{\beta }\in $ $\mathfrak{R}$ of the variational problem (\ref{BCS pressure
2}) is always bounded. In fact, $0\leq \mathit{r}_{\beta }\leq \mathit{r}%
_{\max }\leq 1/4$. Up to (critical) points corresponding to a first order
phase transition it is always unique and continuous w.r.t. each parameter.
For low inverse temperatures $\beta $ (high temperature regime)
straightforward computations show that $\mathit{r}_{\beta }=0$. On the other
hand, for large coupling constants $\gamma _{\mathit{s}}>0$ we have the
existence of a unique strictly positive solution $\mathit{r}_{\beta }>0$. In
this case any non--zero solution $\mathit{r}_{\beta }$ of the variational
problem (\ref{BCS pressure 2}) has to be solution of the gap equation (or
Euler--Lagrange equation):%
\begin{equation}
\mathrm{G}_{0}\left( \mathit{r}_{\beta }\right) +\mathrm{G}_{1}\left(
\mathit{r}_{\beta }\right) +\mathrm{G}_{2}\left( \mathit{r}_{\beta }\right) =%
\frac{2}{\gamma _{\mathit{s}}}  \label{BCS gap equation}
\end{equation}%
with for all $r\geq 0$,%
\begin{eqnarray*}
\mathrm{G}_{0}\left( r\right) &:&=\frac{\sinh (\beta g_{r,0})}{\mathrm{f}(r)%
\mathrm{e}^{\beta \left( \lambda _{\mathit{s}}+\mu _{\mathit{f}}\right)
}g_{r,0}}\ , \\
\mathrm{G}_{1}\left( r\right) &:&=2\frac{\cosh (\beta h_{\mathit{f}})\sinh
(\beta g_{r,2\lambda })}{\mathrm{f}(r)\mathrm{e}^{\beta \left( 2\lambda
+\lambda _{\mathit{s}}\right) }g_{r,2\lambda }}\ , \\
\mathrm{G}_{2}\left( r\right) &:&=\frac{\sinh (\beta g_{r,4\lambda })}{%
\mathrm{f}(r)\mathrm{e}^{\beta (4\lambda +\lambda _{\mathit{s}}+2\lambda _{%
\mathit{f}}-\mu _{\mathit{f}})}g_{r,4\lambda }}\ .
\end{eqnarray*}%
Therefore, the set
\begin{multline}
\mathcal{S}:=\Big\{\mu _{\mathit{s}},h_{\mathit{s}},\mu _{\mathit{f}},h_{%
\mathit{f}},\eta \in \mathbb{R},\ \gamma _{\mathit{s}},\lambda _{\mathit{s}%
},\lambda _{\mathit{f}},\lambda \geq 0,\ \beta >0\;:\;
\label{critical point open set} \\
\qquad \text{The solution }\mathit{r}_{\beta }>0\text{ of (\ref{BCS pressure
2}) is unique}\Big\}  \notag
\end{multline}%
is non--empty. Analogously, the set of parameters
\begin{multline*}
\mathcal{S}_{0}:=\Big\{\mu _{\mathit{s}},h_{\mathit{s}},\mu _{\mathit{f}},h_{%
\mathit{f}},\eta \in \mathbb{R},\ \gamma _{\mathit{s}},\lambda _{\mathit{s}%
},\lambda _{\mathit{f}},\lambda \geq 0,\ \beta >0\;:\; \\
\qquad \mathit{r}_{\beta }=0\text{ is unique solution of (\ref{BCS pressure
2})}\Big\}
\end{multline*}%
is equally not empty. The intersection $\mathcal{C}:=$ $\mathcal{S}^{c}\cap
\mathcal{S}_{0}^{c}$ of the complements $\mathcal{S}^{c}$ and $\mathcal{S}%
_{0}^{c}$ in the set
\begin{equation*}
\Big\{\mu _{\mathit{s}},h_{\mathit{s}},\mu _{\mathit{f}},h_{\mathit{f}},\eta
\in \mathbb{R},\ \gamma _{\mathit{s}},\lambda _{\mathit{s}},\lambda _{%
\mathit{f}},\lambda \geq 0,\ \beta >0\Big\}
\end{equation*}%
of $\mathcal{S}$ and $\mathcal{S}_{0}$ respectively, is called the set of
(first order) critical points of the model. It consists per definition of
all combinations of parameters for which (\ref{BCS pressure 2}) has more
than one solution. In fact, $\mathcal{S}$ corresponds to the purely
superconducting phase since the order parameter solution of (\ref{BCS
pressure 2}) can be interpreted as the Cooper pair condensate density $%
\omega _{N}(\mathfrak{c}_{0}^{\ast }\mathfrak{c}_{0})/N$ as $N\rightarrow
\infty $ where
\begin{equation*}
\mathfrak{c}_{0}:=\frac{1}{\sqrt{N}}\sum_{x\in \Lambda _{N}}a_{x,\downarrow
}^{(\mathit{s})}a_{x,\uparrow }^{(\mathit{s})}=\frac{1}{\sqrt{N}}%
\sum\limits_{k\in \Lambda _{N}^{\ast }}\tilde{a}_{k,\downarrow }^{(\mathit{s}%
)}\tilde{a}_{-k,\uparrow }^{(\mathit{s})}
\end{equation*}%
(resp. $\mathfrak{c}_{0}^{\ast }$) annihilates (resp. creates) one Cooper
pair of \textquotedblleft $\mathit{s}$\textquotedblright -- fermions within
the condensate, i.e., in the zero--mode for\ pairs of \textquotedblleft $%
\mathit{s}$\textquotedblright --fermions. Indeed, away from any critical
point the Cooper pair condensate density equals
\begin{equation}
\underset{N\rightarrow \infty }{\lim }\left\{ N^{-1}\omega _{N}\left(
\mathfrak{c}_{0}^{\ast }\mathfrak{c}_{0}\right) \right\} =\mathit{r}_{\beta
}\leq 1/4.  \label{griffiths1}
\end{equation}%
As already explained, the proof of this last result follows from a rather
explicit description of the weak$^{\ast }$--limit points of local Gibbs
states $\omega _{N}$. We omit details as this fact is a simple adaptation of
the results of \cite{BruPedra1}. See also Section \ref{appendix}. This
superconducting phase is a purely \emph{s--wave} phase with an off--diagonal
long range order as described in \cite[Thm 3.2 \& 3.3]{BruPedra1}.

Observe also that an analysis of the intra--band density correlation%
\begin{equation*}
\underset{N\rightarrow \infty }{\lim }\omega _{N}\left( n_{x,\uparrow }^{(%
\mathit{s})}n_{x,\downarrow }^{(\mathit{s})}\right) \in \lbrack 0,1]\
\end{equation*}%
for \textquotedblleft $\mathit{s}$\textquotedblright --fermions\ allows us
to characterize the difference between the superconducting and
non--superconducting phases in terms of space distributions of
\textquotedblleft $\mathit{s}$\textquotedblright --fermions:

\begin{itemize}
\item[(n)] \emph{Within the non--superconducting phase.} The probability of
finding two \textquotedblleft $\mathit{s}$\textquotedblright --fermions on
the same site goes to zero as $N\rightarrow \infty $ and $\beta \rightarrow
\infty $, whereas the mean density of \textquotedblleft $\mathit{s}$%
\textquotedblright --fermions goes to one. In other words, the probability
of finding exactly one \textquotedblleft $\mathit{s}$\textquotedblright
--fermion on one given site is one in the limit $N\rightarrow \infty $, $%
\beta \rightarrow \infty $. This characterizes an insulating Mott phase and
corresponds to well--known experimental facts related to cuprates outside
the superconducting phase and near half--filling.

\item[(s)] \emph{Within the superconducting phase.} 100\% of
\textquotedblleft $\mathit{s}$\textquotedblright --fermions form Cooper
pairs in the limit of zero--temperature in the sense that the conditional
probability of finding a second \textquotedblleft $\mathit{s}$%
\textquotedblright --fermion in one given site provided there is already one
fermion in that site goes to one in the limit $N\rightarrow \infty $, $\beta
\rightarrow \infty $. A high pairing fraction for carriers is a well--known
property of real high $T_{c}$--materials in contrast to the low pairing
fraction of conventional superconductors.
\end{itemize}

\noindent Such a study is performed in \cite[Sect. 3.4]{BruPedra1} for the
one--band case and the arguments can easily be translated into the two--band
case. We omit hence the details and only remark that the highest Cooper pair
condensate density $\mathit{r}_{\beta }$ is in fact $1/4$. Therefore, even
if all \textquotedblleft $\mathit{s}$\textquotedblright\ -- fermions form
Cooper pairs at small temperatures, there is at most $50\%$ of fermion pairs
in the condensate. In other words there is always a \emph{depletion} of the
condensate exactly as in the one--band case (cf. \cite[Fig. 8]{BruPedra1}).

Other two important thermodynamic functions we need in the present section
are the (infinite volume) densities per lattice site of \textquotedblleft $%
\mathit{s}$\textquotedblright --\ and \textquotedblleft $\mathit{f}$%
\textquotedblright --fermions: \
\begin{eqnarray*}
\mathit{d}_{\beta }^{\left( \mathit{s}\right) } &:&=\underset{N\rightarrow
\infty }{\lim }\omega _{N}\left( n_{x,\uparrow }^{(\mathit{s}%
)}+n_{x,\downarrow }^{(\mathit{s})}\right) \in \lbrack 0,2]\ , \\
\mathit{d}_{\beta }^{\left( \mathit{f}\right) } &:&=\underset{N\rightarrow
\infty }{\lim }\omega _{N}\left( n_{x,\uparrow }^{(\mathit{f}%
)}+n_{x,\downarrow }^{(\mathit{f})}\right) \in \lbrack 0,2]\ .
\end{eqnarray*}%
Using standard computations (cf. Section \ref{appendix}), both densities can
be explicitly computed. They are not depending on $x\in
%TCIMACRO{\U{2124} }%
%BeginExpansion
\mathbb{Z}
%EndExpansion
^{D}$ and respectively equal to%
\begin{eqnarray}
\mathit{d}_{\beta }^{\left( \mathit{s}\right) } &=&1+(\mu _{\mathit{s}%
}-\lambda _{\mathit{s}})\mathrm{G}_{0}\left( \mathit{r}_{\beta }\right)
+(\mu _{\mathit{s}}-2\lambda -\lambda _{\mathit{s}})\mathrm{G}_{1}\left(
\mathit{r}_{\beta }\right)  \notag \\
&&+(\mu _{\mathit{s}}-4\lambda -\lambda _{\mathit{s}})\mathrm{G}_{2}\left(
\mathit{r}_{\beta }\right)  \label{griffiths2}
\end{eqnarray}%
and
\begin{eqnarray}
\mathit{d}_{\beta }^{\left( \mathit{f}\right) } &=&1+\frac{1}{\mathrm{f}(%
\mathit{r}_{\beta })}\Bigl\{(\mathrm{e}^{-\beta (4\lambda +2\lambda _{%
\mathit{f}}-\mu _{\mathit{f}})}-\mathrm{e}^{-\beta \mu _{\mathit{f}}})\cosh
(\beta h_{\mathit{s}})  \notag \\
&&+\mathrm{e}^{-\beta (\lambda _{\mathit{s}}+4\lambda +2\lambda _{\mathit{f}%
}-\mu _{\mathit{f}})}\cosh (\beta g_{\mathit{r}_{\beta },4\lambda })  \notag
\\
&&-\mathrm{e}^{-\beta (\lambda _{\mathit{s}}+\mu _{\mathit{f}})}\cosh (\beta
g_{\mathit{r}_{\beta },0})\Bigr\}  \label{griffiths3}
\end{eqnarray}%
for any $\mu _{\mathit{s}},h_{\mathit{s}},\mu _{\mathit{f}},h_{\mathit{f}%
},\eta \in \mathbb{R}$, $\gamma _{\mathit{s}},\lambda _{\mathit{s}},\lambda
_{\mathit{f}},\lambda \geq 0$ and $\beta >0$ away from any critical point.

Now we are in position to analyze the first case (a) about the effect of an\
almost empty\ \textquotedblleft $\mathit{f}$\textquotedblright --band\ ($%
\delta <0$) on superconductivity. For simplicity we always take $\eta =0$ in
the four cases (a)--(d) in order to avoid complicated cross--effects due to
inter--band magnetic interactions. Indeed, the thermodynamics for $\eta \neq
0$ will be analyzed separately afterwards. \medskip

\textbf{(a) The almost empty \textquotedblleft }$\mathit{f}$\textbf{%
\textquotedblright --band:}\medskip\

\begin{figure}[hbtp]
\includegraphics[angle=0,scale=1,clip=true,width=6.5cm]{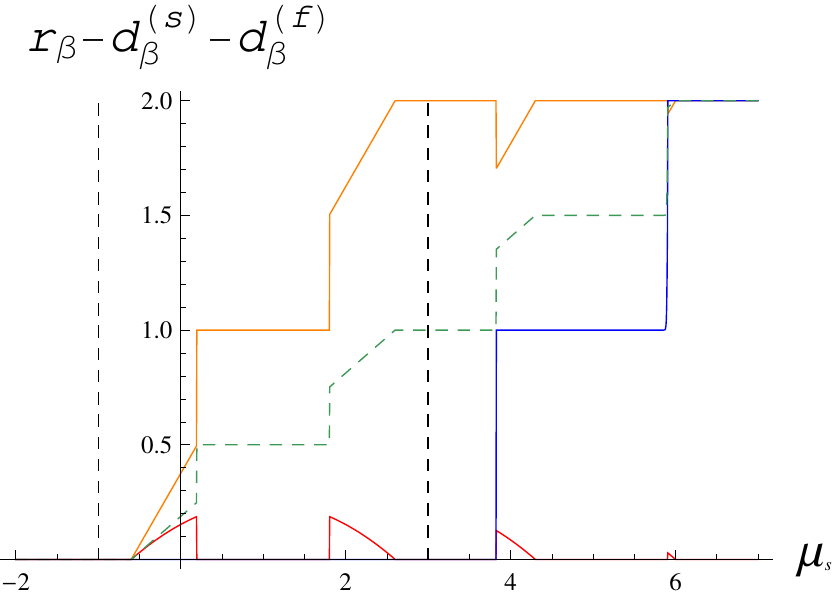}
\caption{\emph{Illustration
of the Cooper pair condensate density $\textit{r}_{\beta }$ (red), the
\textquotedblleft $\textit{s}$\textquotedblright--fermion\  density $\textit{d}_{\beta }^{\left( \textit{s}\right) }$ (orange), and the \textquotedblleft $\textit{f}$\textquotedblright--fermion\ density $\textit{d}_{\beta
}^{\left( \textit{f}\right) }$ (blue) for $\beta=200$, $\gamma _{\textit{s}}=3.2$, $\delta =-0.5$, $\lambda _{\textit{s}}=\lambda _{\textit{f}}=1$, $\lambda =0.85
$, $h_{\textit{s}}=h_{\textit{f}}=\eta=0$, and $\mu _{\textit{s}}\in \left[ -2,7\right] $. The dashed green line corresponds
to the average density $\textit{d}_{\beta }:=(\textit{d}_{\beta }^{\left( \textit{s}\right) }+\textit{d}_{\beta }^{\left( \textit{f}\right) })/2$. We concentrate our
study between both vertical dashed black lines $\mu _{\textit{s}}=-1$ and $\mu _{\textit{s}}=3$.}}
\label{mufixed0bis.eps}
\end{figure}

The first phase diagram to be discussed is given in Fig. \ref%
{mufixed0bis.eps}. For our purposes here, the most interesting interval of
chemical potentials lies between the vertical dashed black lines $\mu _{%
\mathit{s}}=-1$ and $\mu _{\mathit{s}}=3$. The case\ $\beta =200$ is
representative for the low temperature regime $\beta \rightarrow \infty $.
Observe in Fig. \ref{mufixed0bis.eps} the existence for $\mu _{\mathit{s}%
}\in \left[ -1,3\right] $ of a superconductor--Mott insulator phase
transition as described in \cite[Sect. 3.5]{BruPedra1}. Notice, moreover,
that for $\mu _{\mathit{s}}\in \left[ -1,3\right] $ \ (and $\beta >>1$) the
fermion density in the \textquotedblleft $\mathit{f}$\textquotedblright
\textbf{--}band\textbf{\ }is almost zero and one could suppose that the
influence of this band is negligible and the behavior of the system is well
described in this interval of chemical potentials by some effective
one--band model for \textquotedblleft $\mathit{s}$\textquotedblright \textbf{%
--}fermions. \ The usual electron--hole symmetry of phase diagrams of the
one--band model -- as discussed above -- would then suggest that, also in
the two--band case, superconductivity below half--filling ($\mathit{d}%
_{\beta }^{\left( \mathit{s}\right) }<1$) is as favorable as
superconductivity above half--filling ($\mathit{d}_{\beta }^{\left( \mathit{s%
}\right) }>1$) -- at least for $\mu _{\mathit{s}}\in \left[ -1,3\right] $.
%TCIMACRO{%
%\TeXButton{mufixed0bis.eps}{\begin{figure}[hbtp]
%\includegraphics[angle=0,scale=1,clip=true,width=6.5cm]{mufixed0bis}
%\caption{\emph{Illustration
%of the Cooper pair condensate density $\textit{r}_{\beta }$ (red), the
%\textquotedblleft $\textit{s}$\textquotedblright--fermion\  density $\textit{d}_{\beta }^{\left( \textit{s}\right) }$ (orange), and the \textquotedblleft $\textit{f}$\textquotedblright--fermion\ density $\textit{d}_{\beta
%}^{\left( \textit{f}\right) }$ (blue) for $\beta=200$, $\gamma _{\textit{s}}=3.2$, $\delta =-0.5$, $\lambda _{\textit{s}}=\lambda _{\textit{f}}=1$, $\lambda =0.85
%$, $h_{\textit{s}}=h_{\textit{f}}=\eta=0$, and $\mu _{\textit{s}}\in \left[ -2,7\right] $. The dashed green line corresponds
%to the average density $\textit{d}_{\beta }:=(\textit{d}_{\beta }^{\left( \textit{s}\right) }+\textit{d}_{\beta }^{\left( \textit{f}\right) })/2$. We concentrate our
%study between both vertical dashed black lines $\mu _{\textit{s}}=-1$ and $\mu _{\textit{s}}=3$.}}
%\label{mufixed0bis.eps}
%\end{figure}}}%
%BeginExpansion
%EndExpansion
%TCIMACRO{%
%\TeXButton{magncritical0d0bis.eps}{\begin{figure}[hbtp]
%\includegraphics[angle=0,scale=1,clip=true,width=6.5cm]{magncritical0d0bis}
%\caption{\emph{Illustration of the critical temperature $\theta _{c}$ for $\gamma _{\textit{s}}=3.2$, $\delta =-0.5$, $\lambda _{\textit{s}}=\lambda _{\textit{f}}=1$, $\lambda =0.85
%$, $h_{\textit{s}}=h_{\textit{f}}=\eta=0$, and $\mu _{\textit{s}}\in \left[ -1,3\right] $.
%}}
%\label{magncritical0d0bis.eps}
%\end{figure}}}%
%BeginExpansion
\begin{figure}[hbtp]
\includegraphics[angle=0,scale=1,clip=true,width=6.5cm]{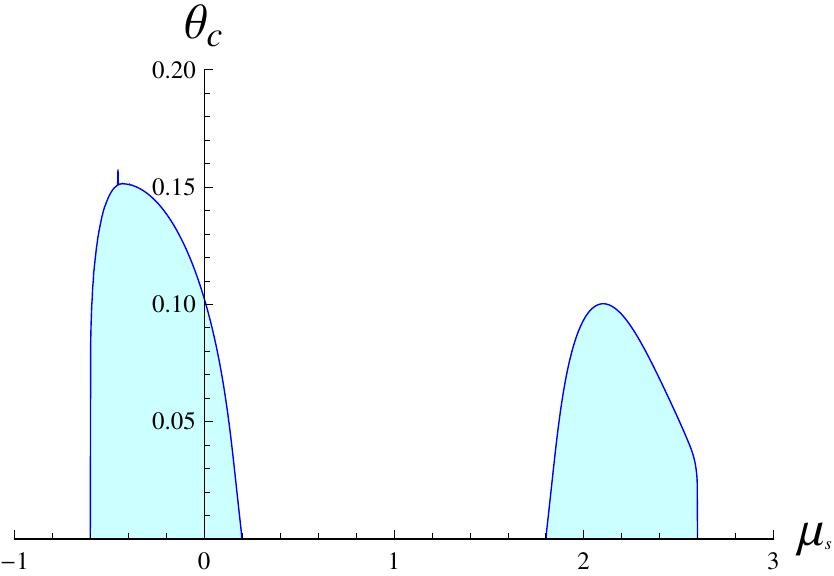}
\caption{\emph{Illustration of the critical temperature $\theta _{c}$ for $\gamma _{\textit{s}}=3.2$, $\delta =-0.5$, $\lambda _{\textit{s}}=\lambda _{\textit{f}}=1$, $\lambda =0.85
$, $h_{\textit{s}}=h_{\textit{f}}=\eta=0$, and $\mu _{\textit{s}}\in \left[ -1,3\right] $.
}}
\label{magncritical0d0bis.eps}
\end{figure}%
%EndExpansion

\begin{figure}[hbtp]
\includegraphics[angle=0,scale=1,clip=true,width=6.5cm]{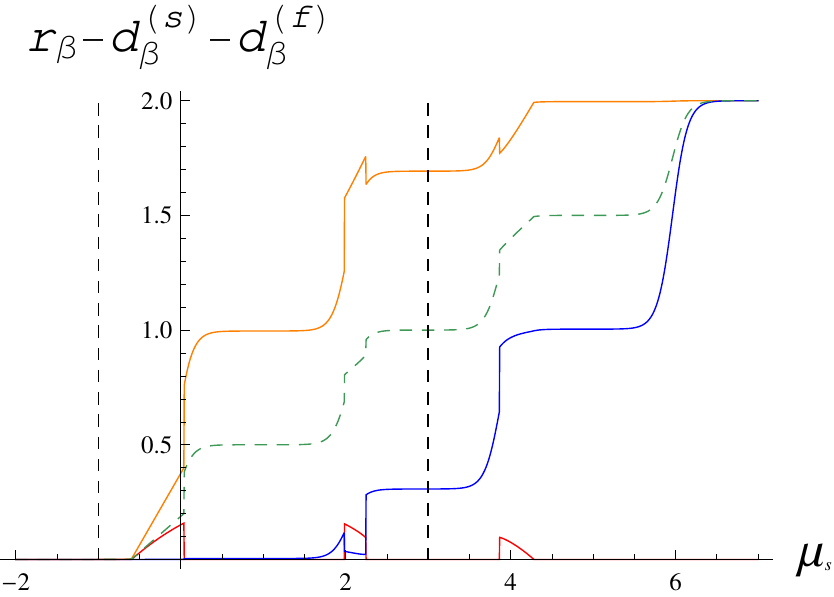}
\caption{\emph{Illustration of the Cooper pair condensate density $\textit{r}_{\beta }$ (red), the
\textquotedblleft $\textit{s}$\textquotedblright--fermion\ density $\textit{d}_{\beta }^{\left( \textit{s}\right) }$ (orange), and the
\textquotedblleft $\textit{f}$\textquotedblright--fermion\ density $\textit{d}_{\beta
}^{\left( \textit{f}\right) }$ (blue) for $\beta=11$, $\gamma _{\textit{s}}=3.2$, $\delta =-0.5$, $\lambda _{\textit{s}}=\lambda _{\textit{f}}=1$, $\lambda =0.85
$, $h_{\textit{s}}=h_{\textit{f}}=\eta=0$, and $\mu _{\textit{s}}\in \left[ -2,7\right] $. The dashed green line corresponds
to the average density $\textit{d}_{\beta }:=(\textit{d}_{\beta }^{\left( \textit{s}\right) }+\textit{d}_{\beta }^{\left( \textit{f}\right) })/2$. We concentrate our
study between both vertical dashed black lines $\mu _{\textit{s}}=-1$ and $\mu _{\textit{s}}=3$.}}
\label{mufixed0.eps}
\end{figure}

\begin{figure}[hbtp]
\includegraphics[angle=0,scale=1,clip=true,width=6.5cm]{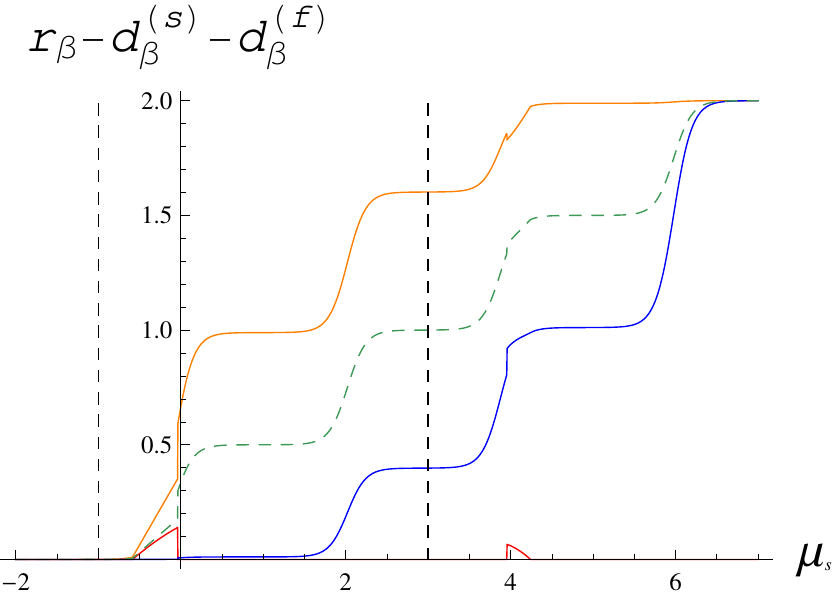}
\caption{\emph{Illustration of the Cooper pair condensate density $\textit{r}_{\beta }$ (red), the
\textquotedblleft $\textit{s}$\textquotedblright--fermion\ density $\textit{d}_{\beta }^{\left( \textit{s}\right) }$ (orange), and the
\textquotedblleft $\textit{f}$\textquotedblright--fermion\ density $\textit{d}_{\beta
}^{\left( \textit{f}\right) }$ (blue) for $\beta=9$, $\gamma _{\textit{s}}=3.2$, $\delta =-0.5$, $\lambda _{\textit{s}}=\lambda _{\textit{f}}=1$, $\lambda =0.85
$, $h_{\textit{s}}=h_{\textit{f}}=\eta=0$, and $\mu _{\textit{s}}\in \left[ -2,7\right] $. The dashed green line corresponds
to the average density $\textit{d}_{\beta }:=(\textit{d}_{\beta }^{\left( \textit{s}\right) }+\textit{d}_{\beta }^{\left( \textit{f}\right) })/2$. We concentrate our
study between both vertical dashed black lines $\mu _{\textit{s}}=-1$ and $\mu _{\textit{s}}=3$.}}
\label{mufixed0bisbisbis.eps}
\end{figure}

Nevertheless, this prediction is \emph{not correct.} Indeed, for $\mu _{%
\mathrm{s}}\in \left[ -1,3\right] $ the critical temperatures $\theta _{c}$
of superconductivity for $\mathit{d}_{\beta }^{\left( \mathit{s}\right) }>1$
are generally \emph{lower} than the ones for $\mathit{d}_{\beta }^{\left(
\mathit{s}\right) }<1$, see Fig. \ref{magncritical0d0bis.eps}. This
asymmetry below and above half--filling results from the (screened) Coulomb
interaction (\ref{coulomb repulsionbis}) together with thermal excitations
of \textquotedblleft $\mathit{s}$\textquotedblright --fermions\ into the --
here energetically higher ($\delta <0$) -- \textquotedblleft $\mathit{f}$%
\textquotedblright --band as one can see from Fig. \ref{mufixed0.eps} (which
represents the case of an inverse temperature $\beta =11$). In fact, for
higher temperatures (lower $\beta $) and $\mathit{d}_{\beta }^{\left(
\mathit{s}\right) }>1$ the superconducting phase completely disappears,
whereas for $\mathit{d}_{\beta }^{\left( \mathit{s}\right) }<1$, the $U(1)$%
--broken phase is still found. See Fig. \ref{mufixed0bisbisbis.eps} which
corresponds to the case $\beta =9$.%
%TCIMACRO{%
%\TeXButton{mufixed0.eps}{\begin{figure}[hbtp]
%\includegraphics[angle=0,scale=1,clip=true,width=6.5cm]{mufixed0}
%\caption{\emph{Illustration of the Cooper pair condensate density $\textit{r}_{\beta }$ (red), the
%\textquotedblleft $\textit{s}$\textquotedblright--fermion\ density $\textit{d}_{\beta }^{\left( \textit{s}\right) }$ (orange), and the
%\textquotedblleft $\textit{f}$\textquotedblright--fermion\ density $\textit{d}_{\beta
%}^{\left( \textit{f}\right) }$ (blue) for $\beta=11$, $\gamma _{\textit{s}}=3.2$, $\delta =-0.5$, $\lambda _{\textit{s}}=\lambda _{\textit{f}}=1$, $\lambda =0.85
%$, $h_{\textit{s}}=h_{\textit{f}}=\eta=0$, and $\mu _{\textit{s}}\in \left[ -2,7\right] $. The dashed green line corresponds
%to the average density $\textit{d}_{\beta }:=(\textit{d}_{\beta }^{\left( \textit{s}\right) }+\textit{d}_{\beta }^{\left( \textit{f}\right) })/2$. We concentrate our
%study between both vertical dashed black lines $\mu _{\textit{s}}=-1$ and $\mu _{\textit{s}}=3$.}}
%\label{mufixed0.eps}
%\end{figure}}}%
%BeginExpansion
%EndExpansion
%TCIMACRO{%
%\TeXButton{mufixed0bisbisbis.eps}{\begin{figure}[hbtp]
%\includegraphics[angle=0,scale=1,clip=true,width=6.5cm]{mufixed0bisbisbis}
%\caption{\emph{Illustration of the Cooper pair condensate density $\textit{r}_{\beta }$ (red), the
%\textquotedblleft $\textit{s}$\textquotedblright--fermion\ density $\textit{d}_{\beta }^{\left( \textit{s}\right) }$ (orange), and the
%\textquotedblleft $\textit{f}$\textquotedblright--fermion\ density $\textit{d}_{\beta
%}^{\left( \textit{f}\right) }$ (blue) for $\beta=9$, $\gamma _{\textit{s}}=3.2$, $\delta =-0.5$, $\lambda _{\textit{s}}=\lambda _{\textit{f}}=1$, $\lambda =0.85
%$, $h_{\textit{s}}=h_{\textit{f}}=\eta=0$, and $\mu _{\textit{s}}\in \left[ -2,7\right] $. The dashed green line corresponds
%to the average density $\textit{d}_{\beta }:=(\textit{d}_{\beta }^{\left( \textit{s}\right) }+\textit{d}_{\beta }^{\left( \textit{f}\right) })/2$. We concentrate our
%study between both vertical dashed black lines $\mu _{\textit{s}}=-1$ and $\mu _{\textit{s}}=3$.}}
%\label{mufixed0bisbisbis.eps}
%\end{figure}}}%
%BeginExpansion
%EndExpansion

\begin{figure}[hbtp]
\includegraphics[angle=0,scale=1,clip=true,width=6.5cm]{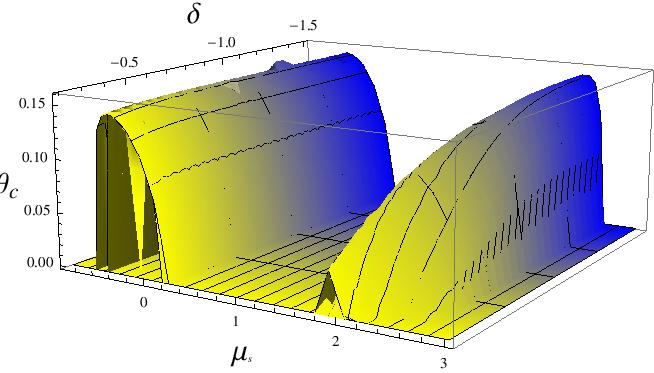}
\caption{\emph{Illustration
of the critical temperature $\theta _{c}$ for $\gamma _{\textit{s}}=3.2$,  $\lambda _{\textit{s}}=\lambda _{\textit{f}}=1$, $\lambda =0.85
$, $h_{\textit{s}}=h_{\textit{f}}=\eta=0$, $\mu _{\textit{s}}\in \left[ -1,3\right] $, and $\delta \in \left[ -0.2,-1.5\right] $. The color from yellow to blue reflects the increase of the
energy gap $|\delta|$.
}}
\label{plot3dgapbis.eps}
\end{figure}

One clearly sees in Figs. \ref{mufixed0.eps} and \ref{mufixed0bisbisbis.eps}
that at large enough temperature the \textquotedblleft $\mathit{f}$%
\textquotedblright --band\ is not anymore empty and can inhibit the
formation of Cooper pairs in the \textquotedblleft $\mathit{s}$%
\textquotedblright --band because of the Coulomb repulsion (\ref{coulomb
repulsionbis}). This phenomenon can only appear if the critical temperature $%
\theta _{c}$ is large enough or/and the energy difference $\left\vert \delta
\right\vert $ is small enough in order to get a non--negligible population
of fermions in the \textquotedblleft $\mathit{f}$\textquotedblright --band\
through thermal fluctuations at $\theta \lessapprox \theta _{c}$. Indeed,
for large enough $\left\vert \delta \right\vert $ the electron--hole
asymmetry disappears, i.e., a critical value of the energy gap $\left\vert
\delta \right\vert $ for electron--hole asymmetry seems to exist, see Fig. %
\ref{plot3dgapbis.eps}. If $\delta <0$ and the \textquotedblleft $\mathit{f}$%
\textquotedblright \textbf{--}band is empty at zero temperature then -- by a
huge amount of numerical experiments -- \emph{no} choice of parameters can
favor superconductivity above half--filling, i.e., the maximal critical
temperature $\theta _{c}$ is always attained at $\mathit{d}_{\beta }^{\left(
\mathit{s}\right) }<1$, as expected.
%TCIMACRO{%
%\TeXButton{plot3dgapbis.eps}{\begin{figure}[hbtp]
%\includegraphics[angle=0,scale=1,clip=true,width=6.5cm]{plot3dgapbis}
%\caption{\emph{Illustration
%of the critical temperature $\theta _{c}$ for $\gamma _{\textit{s}}=3.2$,  $\lambda _{\textit{s}}=\lambda _{\textit{f}}=1$, $\lambda =0.85
%$, $h_{\textit{s}}=h_{\textit{f}}=\eta=0$, $\mu _{\textit{s}}\in \left[ -1,3\right] $, and $\delta \in \left[ -0.2,-1.5\right] $. The color from yellow to blue reflects the increase of the
%energy gap $|\delta|$.
%}}
%\label{plot3dgapbis.eps}
%\end{figure}}}%
%BeginExpansion
%EndExpansion

Observe that the total density of fermions (and not necessarily the chemical
potential $\mu _{\mathit{s}}$) should be considered as being a fixed
quantity in models for high--$T_{c}$\emph{\ }cuprate\emph{\ }%
superconductors. It is therefore important to check whether the phenomenon
described above at fixed chemical potential $\mu _{\mathit{s}}$ is also seen
w.r.t. fixed total densities $\rho \in (0,2)$ of fermions or not. Indeed, by
strict convexity of the pressure at any finite volume the total density of
fermions
\begin{equation}
\mathit{d}_{\beta }:=\frac{1}{2}(\mathit{d}_{\beta }^{\left( \mathit{s}%
\right) }+\mathit{d}_{\beta }^{\left( \mathit{f}\right) })\in (0,2)
\label{full density}
\end{equation}%
per lattice site and per band is strictly increasing as a function of the
chemical potential $\mu _{\mathit{s}}$. Therefore, for any fixed $h_{\mathit{%
s}},\delta ,h_{\mathit{f}},\eta \in \mathbb{R}$, $\gamma _{\mathit{s}%
},\lambda _{\mathit{s}},\lambda _{\mathit{f}},\lambda \geq 0$, $\beta >0$
and $\rho \in (0,2)$ there exists a unique real number $\mu _{\mathit{s}%
,N,\beta }$ such that
\begin{eqnarray}
\rho &=&\dfrac{1}{2N}\sum_{x\in \Lambda _{N}}\omega _{N}\left( n_{x,\uparrow
}^{(\mathit{s})}+n_{x,\downarrow }^{(\mathit{s})}+n_{x,\uparrow }^{(\mathit{f%
})}+n_{x,\downarrow }^{(\mathit{f})}\right)
\label{mu fixed particle density} \\
&=&\dfrac{1}{2}\omega _{N}\left( n_{x,\uparrow }^{(\mathit{s}%
)}+n_{x,\downarrow }^{(\mathit{s})}+n_{x,\uparrow }^{(\mathit{f}%
)}+n_{x,\downarrow }^{(\mathit{f})}\right) ,\quad \forall x\in \Lambda _{N},
\notag
\end{eqnarray}%
where $\omega _{N}$ represents the (finite volume) grand--canonical Gibbs
state (\ref{BCS gibbs state Hn}) associated with $\mathrm{H}_{N}^{(\mathit{s}%
,\mathit{f})}$ and taken at inverse temperature $\beta >0$ and chemical
potentials $\mu _{\mathit{s},N,\beta }$ and $\mu _{\mathit{f}}=\mu _{\mathit{%
s},N,\beta }+\delta $. The Gibbs state $\omega _{N}$ can also in this case
be explicitly computed in the limit $N\rightarrow \infty $, see \cite[Thm 6.5%
]{BruPedra1}.

\begin{figure}[hbtp]
\includegraphics[angle=0,scale=1,clip=true,width=6.5cm]{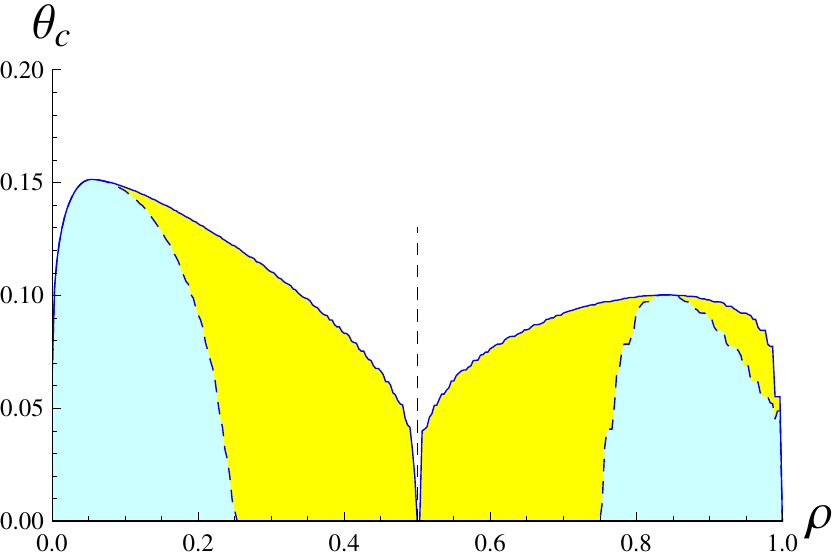}
\caption{\emph{Illustration
of the critical temperature $\theta _{c}$ for $\gamma _{\textit{s}}=3.2$, $\delta =-0.5$, $\lambda _{\textit{s}}=\lambda _{\textit{f}}=1$, $\lambda =0.85
$, $h_{\textit{s}}=h_{\textit{f}}=\eta=0$, and $\rho \in \left[ 0,1\right] $. The blue and yellow region correspond respectively
to the superconducting and ferromagnetic--superconducting phases.
The dashed line corresponds to the absence of doping.
}}
\label{magncritical0d0bisD.eps}
\end{figure}

By representing diagrams w.r.t. the parameter $\rho \in (0,1)$\ (instead of $%
\mu _{\mathit{s}}\in \left[ -1,3\right] )$ we obtain the same kind of
behavior as above, see Fig. \ref{magncritical0d0bisD.eps}. The only
difference to be noted w.r.t. what was already said is the rather frequent
coexistence of superconducting and non--superconducting phases at fixed
total fermion\ densities $\rho $ around $0.5$. Indeed, such a coexistence
takes place because the chemical potentials to be chosen in order to
implement the given densities $\rho $ are such that the parameter vector
lies in the set $\mathcal{C}$ of critical points, see Fig. \ref%
{mufixed0bis.eps}. The mathematical proof of coexistence of phases results
from a detailed analysis of the (weak$^{\ast }$--) limit point of $\omega
_{N}$ which can be performed exactly as done for the one--band case in \cite[%
Thm. 6.5 (ii)]{BruPedra1}.\medskip
%TCIMACRO{%
%\TeXButton{magncritical0d0bisD.eps}{\begin{figure}[hbtp]
%\includegraphics[angle=0,scale=1,clip=true,width=6.5cm]{magncritical0d0bisD}
%\caption{\emph{Illustration
%of the critical temperature $\theta _{c}$ for $\gamma _{\textit{s}}=3.2$, $\delta =-0.5$, $\lambda _{\textit{s}}=\lambda _{\textit{f}}=1$, $\lambda =0.85
%$, $h_{\textit{s}}=h_{\textit{f}}=\eta=0$, and $\rho \in \left[ 0,1\right] $. The blue and yellow region correspond respectively
%to the superconducting and ferromagnetic--superconducting phases.
%The dashed line corresponds to the absence of doping.
%}}
%\label{magncritical0d0bisD.eps}
%\end{figure}}}%
%BeginExpansion
%EndExpansion

\textbf{(b) Breakdown of the half--filled \textquotedblleft }$\mathit{f}$%
\textbf{\textquotedblright --band destroying superconductivity:}\medskip

\begin{figure}[hbtp]
\includegraphics[angle=0,scale=1,clip=true,width=6.5cm]{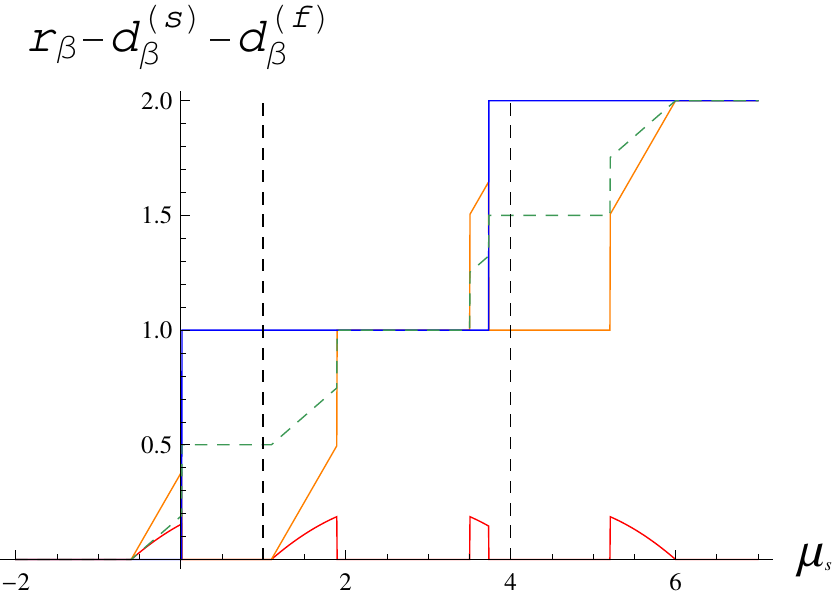}
\caption{\emph{Illustration
of the Cooper pair condensate density $\textit{r}_{\beta }$ (red), the
\textquotedblleft $\textit{s}$\textquotedblright--fermion\ density $\textit{d}_{\beta }^{\left( \textit{s}\right) }$ (orange), and the
\textquotedblleft $\textit{f}$\textquotedblright--fermion\ density $\textit{d}_{\beta
}^{\left( \textit{f}\right) }$ (blue) for $\beta=200$, $\gamma _{\textit{s}}=3.2$, $\delta =0.1$, $\lambda _{\textit{s}}=\lambda _{\textit{f}}=1$, $\lambda =0.85
$, $h_{\textit{s}}=h_{\textit{f}}=\eta=0$, and $\mu _{\textit{s}}\in \left[ -2,7\right] $. The dashed green line corresponds
to the average density $\textit{d}_{\beta }:=(\textit{d}_{\beta }^{\left( \textit{s}\right) }+\textit{d}_{\beta }^{\left( \textit{f}\right) })/2$. We concentrate our
study between both vertical dashed black lines $\mu _{\textit{s}}=1$ and $\mu _{\textit{s}}=4$.}}
\label{mufixed1.eps}
\end{figure}

\begin{figure}[hbtp]
\includegraphics[angle=0,scale=1,clip=true,width=6.5cm]{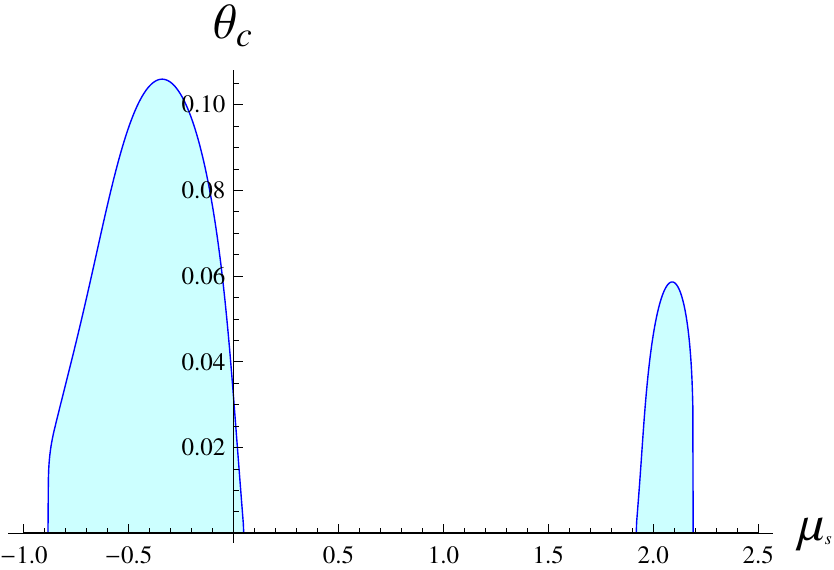}
\caption{\emph{Illustration
of the critical temperature $\theta _{c}$ for $\gamma _{\textit{s}}=3.2$, $\delta =0.1$, $\lambda _{\textit{s}}=\lambda _{\textit{f}}=1$, $\lambda =0.85
$, $h_{\textit{s}}=h_{\textit{f}}=\eta=0$, and $\mu _{\textit{s}}\in \left[ 1,4\right] $.
}}
\label{magncritical0d1.eps}
\end{figure}

This situation corresponds to Fig. \ref{mufixed1.eps}. The interesting
interval of chemical potentials is indicated by the dashed black lines $\mu
_{\mathit{s}}=1$ and $\mu _{\mathit{s}}=4$. All parameters but the gap $%
\delta $ are the same as in Fig. \ref{mufixed0bis.eps}. Indeed, $\delta
=-0.5 $ in case (a), whereas $\delta =0.1$ in case (b). With this last
choice of parameters the fermion density in the \textquotedblleft $\mathit{f}
$\textquotedblright \textbf{--}band is now one (instead of zero) at low
temperatures. The same phenomenon of asymmetry appears as in case (a) for
exactly the same reasons, see Figs. \ref{magncritical0d1.eps} and \ref%
{mufixed1bis.eps}. As in case (a) the maximal critical temperatures for
superconductivity are attained at $\mathit{d}_{\beta }^{\left( \mathit{s}%
\right) }<1$: Superconductivity below half--filling is more favored than
superconductivity above half--filling. The \emph{converse} situation can
also appear: At half--filling of the \textquotedblleft $\mathit{f}$%
\textquotedblright \textbf{--}band, maximal critical temperatures are
attained at $\mathit{d}_{\beta }^{\left( \mathit{s}\right) }>1$ for other
choices of parameters having $\delta <0$ small enough. We omit the details.
With the present choice of parameters observe that increasing the mean
particle density per site and band over the value $\mathit{d}_{\beta }=1$
causes the collapse -- at the same time -- of the superconducting phase in
the \textquotedblleft $\mathit{s}$\textquotedblright \textbf{--}band and of
the half--filling in the \textquotedblleft $\mathit{f}$\textquotedblright
\textbf{--}band, see Fig. \ref{mufixed1.eps}.\bigskip
%TCIMACRO{%
%\TeXButton{mufixed1.eps}{\begin{figure}[hbtp]
%\includegraphics[angle=0,scale=1,clip=true,width=6.5cm]{mufixed1}
%\caption{\emph{Illustration
%of the Cooper pair condensate density $\textit{r}_{\beta }$ (red), the
%\textquotedblleft $\textit{s}$\textquotedblright--fermion\ density $\textit{d}_{\beta }^{\left( \textit{s}\right) }$ (orange), and the
%\textquotedblleft $\textit{f}$\textquotedblright--fermion\ density $\textit{d}_{\beta
%}^{\left( \textit{f}\right) }$ (blue) for $\beta=200$, $\gamma _{\textit{s}}=3.2$, $\delta =0.1$, $\lambda _{\textit{s}}=\lambda _{\textit{f}}=1$, $\lambda =0.85
%$, $h_{\textit{s}}=h_{\textit{f}}=\eta=0$, and $\mu _{\textit{s}}\in \left[ -2,7\right] $. The dashed green line corresponds
%to the average density $\textit{d}_{\beta }:=(\textit{d}_{\beta }^{\left( \textit{s}\right) }+\textit{d}_{\beta }^{\left( \textit{f}\right) })/2$. We concentrate our
%study between both vertical dashed black lines $\mu _{\textit{s}}=1$ and $\mu _{\textit{s}}=4$.}}
%\label{mufixed1.eps}
%\end{figure}}}%
%BeginExpansion
%EndExpansion
%TCIMACRO{%
%\TeXButton{magncritical0d1.eps}{\begin{figure}[hbtp]
%\includegraphics[angle=0,scale=1,clip=true,width=6.5cm]{magncritical0d1}
%\caption{\emph{Illustration
%of the critical temperature $\theta _{c}$ for $\gamma _{\textit{s}}=3.2$, $\delta =0.1$, $\lambda _{\textit{s}}=\lambda _{\textit{f}}=1$, $\lambda =0.85
%$, $h_{\textit{s}}=h_{\textit{f}}=\eta=0$, and $\mu _{\textit{s}}\in \left[ 1,4\right] $.
%}}
%\label{magncritical0d1.eps}
%\end{figure}}}%
%BeginExpansion
%EndExpansion
%TCIMACRO{%
%\TeXButton{mufixed1bis.eps}{\begin{figure}[hbtp]
%\includegraphics[angle=0,scale=1,clip=true,width=6.5cm]{mufixed1bis}
%\caption{\emph{Illustration
%of the Cooper pair condensate density $\textit{r}_{\beta }$ (red), the
%\textquotedblleft $\textit{s}$\textquotedblright--fermion\ density $\textit{d}_{\beta }^{\left( \textit{s}\right) }$ (orange), and the
%\textquotedblleft $\textit{f}$\textquotedblright--fermion\ density $\textit{d}_{\beta
%}^{\left( \textit{f}\right) }$ (blue) for $\beta=17.4$, $\gamma _{\textit{s}}=3.2$, $\delta =0.1$, $\lambda _{\textit{s}}=\lambda _{\textit{f}}=1$, $\lambda =0.85
%$, $h_{\textit{s}}=h_{\textit{f}}=\eta=0$, and $\mu _{\textit{s}}\in \left[ -2,7\right] $. The dashed green line corresponds
%to the average density $\textit{d}_{\beta }:=(\textit{d}_{\beta }^{\left( \textit{s}\right) }+\textit{d}_{\beta }^{\left( \textit{f}\right) })/2$. We concentrate our
%study between both vertical dashed black lines $\mu _{\textit{s}}=1$ and $\mu _{\textit{s}}=4$.}}
%\label{mufixed1bis.eps}
%\end{figure}}}%
%BeginExpansion
\begin{figure}[hbtp]
\includegraphics[angle=0,scale=1,clip=true,width=6.5cm]{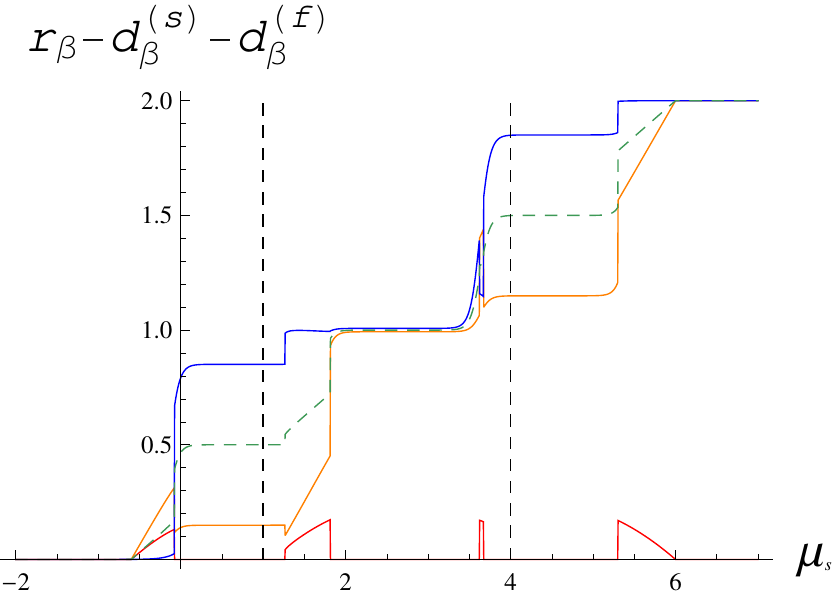}
\caption{\emph{Illustration
of the Cooper pair condensate density $\textit{r}_{\beta }$ (red), the
\textquotedblleft $\textit{s}$\textquotedblright--fermion\ density $\textit{d}_{\beta }^{\left( \textit{s}\right) }$ (orange), and the
\textquotedblleft $\textit{f}$\textquotedblright--fermion\ density $\textit{d}_{\beta
}^{\left( \textit{f}\right) }$ (blue) for $\beta=17.4$, $\gamma _{\textit{s}}=3.2$, $\delta =0.1$, $\lambda _{\textit{s}}=\lambda _{\textit{f}}=1$, $\lambda =0.85
$, $h_{\textit{s}}=h_{\textit{f}}=\eta=0$, and $\mu _{\textit{s}}\in \left[ -2,7\right] $. The dashed green line corresponds
to the average density $\textit{d}_{\beta }:=(\textit{d}_{\beta }^{\left( \textit{s}\right) }+\textit{d}_{\beta }^{\left( \textit{f}\right) })/2$. We concentrate our
study between both vertical dashed black lines $\mu _{\textit{s}}=1$ and $\mu _{\textit{s}}=4$.}}
\label{mufixed1bis.eps}
\end{figure}%
%EndExpansion

\textbf{(c) Breakdown of the half--filled \textquotedblleft }$\mathit{f}$%
\textbf{\textquotedblright --band implying superconductivity:}\medskip

\begin{figure}[hbtp]
\includegraphics[angle=0,scale=1,clip=true,width=6.5cm]{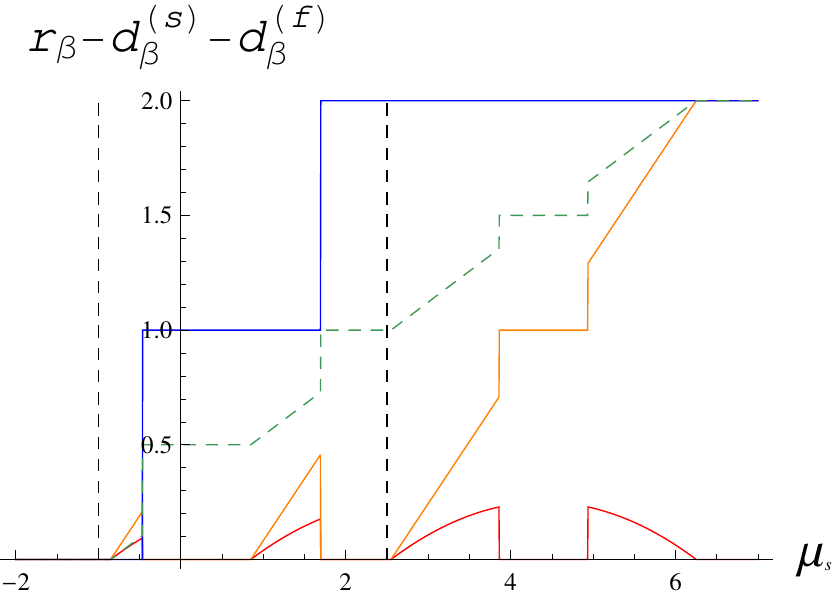}
\caption{\emph{Illustration
of the Cooper pair condensate density $\textit{r}_{\beta }$ (red), the
\textquotedblleft $\textit{s}$\textquotedblright--fermion\ density $\textit{d}_{\beta }^{\left( \textit{s}\right) }$ (orange), and the
\textquotedblleft $\textit{f}$\textquotedblright--fermion\ density $\textit{d}_{\beta
}^{\left( \textit{f}\right) }$ (blue) for $\beta=200$, $\gamma _{\textit{s}}=3.7$, $\delta =0.5$, $\lambda _{\textit{s}}=\lambda _{\textit{f}}=1$, $\lambda =0.85
$, $h_{\textit{s}}=h_{\textit{f}}=\eta=0$, and $\mu _{\textit{s}}\in \left[ -2,7\right] $. The dashed green line corresponds
to the average density $\textit{d}_{\beta }:=(\textit{d}_{\beta }^{\left( \textit{s}\right) }+\textit{d}_{\beta }^{\left( \textit{f}\right) })/2$. We concentrate our
study between both vertical dashed black lines $\mu _{\textit{s}}=-1$ and $\mu _{\textit{s}}=2.5$.}}
\label{mufixed105.eps}
\end{figure}

\begin{figure}[hbtp]
\includegraphics[angle=0,scale=1,clip=true,width=6.5cm]{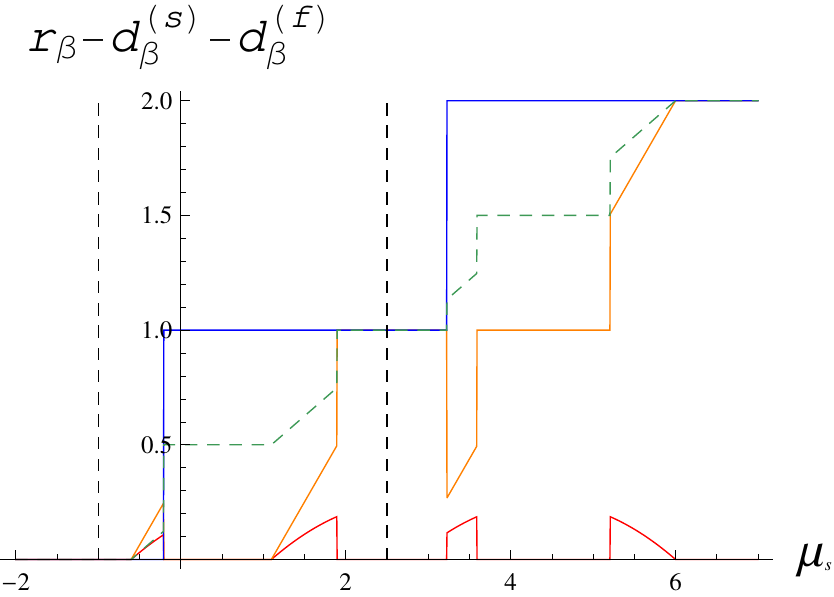}
\caption{\emph{Illustration
of the Cooper pair condensate density $\textit{r}_{\beta }$ (red), the
\textquotedblleft $\textit{s}$\textquotedblright--fermion\ density $\textit{d}_{\beta }^{\left( \textit{s}\right) }$ (orange) and the
\textquotedblleft $\textit{f}$\textquotedblright--fermion\ density $\textit{d}_{\beta
}^{\left( \textit{f}\right) }$ (blue) for $\beta=200$, $\gamma _{\textit{s}}=3.2$, $\delta =0.25$, $\lambda _{\textit{s}}=\lambda _{\textit{f}}=1$, $\lambda =0.85
$, $h_{\textit{s}}=h_{\textit{f}}=\eta=0$, and $\mu _{\textit{s}}\in \left[ -2,7\right] $. The dashed green line corresponds
to the average density $\textit{d}_{\beta }:=(\textit{d}_{\beta }^{\left( \textit{s}\right) }+\textit{d}_{\beta }^{\left( \textit{f}\right) })/2$. We concentrate our
study between both vertical dashed black lines $\mu _{\textit{s}}=-1$ and $\mu _{\textit{s}}=2.5$.}}
\label{mufixed1025.eps}
\end{figure}

In this case we analyze the opposite situation to (b): The fact that the
breakdown of the half--filling in the \textquotedblleft $\mathit{f}$%
\textquotedblright \textbf{--}band can drive the \textquotedblleft $\mathit{s%
}$\textquotedblright \textbf{--}band into a superconducting phase. This is
illustrated in Fig. \ref{mufixed105.eps} for $\delta =0.5$ and $\lambda _{%
\mathit{f}}=1$. It concerns the sector of chemical potentials $\mu _{\mathit{%
s}}$ for which $\mathit{d}_{\beta }\leq 0.5$, $\mathit{d}_{\beta }$ being
the mean particle density (\ref{full density}) per site and per band. This
choice of parameters is motivated by the fact that $\lambda _{\mathrm{Cu}%
}\approx 4\rm{eV}$ and $\delta _{\mathrm{Cu-O}}\leq 0.5\lambda _{\mathrm{Cu%
}}$ in $\mathrm{CuO}_{2}$ layers of cuprates, see discussions at the
beginning of Section \ref{section II}. The most interesting interval of
chemical potentials lies between the vertical dashed black lines $\mu _{%
\mathit{s}}=-1$ and $\mu _{\mathit{s}}=2.5$. This phenomenon is also clearly
represented in Fig \ref{mufixed1025.eps} for $\mathit{d}_{\beta }\leq 0.5$
and a different choice of parameters.
%TCIMACRO{%
%\TeXButton{mufixed105.eps}{\begin{figure}[hbtp]
%\includegraphics[angle=0,scale=1,clip=true,width=6.5cm]{mufixed105}
%\caption{\emph{Illustration
%of the Cooper pair condensate density $\textit{r}_{\beta }$ (red), the
%\textquotedblleft $\textit{s}$\textquotedblright--fermion\ density $\textit{d}_{\beta }^{\left( \textit{s}\right) }$ (orange), and the
%\textquotedblleft $\textit{f}$\textquotedblright--fermion\ density $\textit{d}_{\beta
%}^{\left( \textit{f}\right) }$ (blue) for $\beta=200$, $\gamma _{\textit{s}}=3.7$, $\delta =0.5$, $\lambda _{\textit{s}}=\lambda _{\textit{f}}=1$, $\lambda =0.85
%$, $h_{\textit{s}}=h_{\textit{f}}=\eta=0$, and $\mu _{\textit{s}}\in \left[ -2,7\right] $. The dashed green line corresponds
%to the average density $\textit{d}_{\beta }:=(\textit{d}_{\beta }^{\left( \textit{s}\right) }+\textit{d}_{\beta }^{\left( \textit{f}\right) })/2$. We concentrate our
%study between both vertical dashed black lines $\mu _{\textit{s}}=-1$ and $\mu _{\textit{s}}=2.5$.}}
%\label{mufixed105.eps}
%\end{figure}}}%
%BeginExpansion
%EndExpansion

In the case $\delta =0.5$, $\lambda _{\mathit{f}}=1$ (Fig. \ref%
{mufixed105.eps}) and without any doping, the total density of fermions is
one half, i.e., $\mathit{d}_{\beta }=0.5$, and the densities per lattice
site of \textquotedblleft $\mathit{s}$\textquotedblright --\ and
\textquotedblleft $\mathit{f}$\textquotedblright --fermions are respectively
$\mathit{d}_{\beta }^{\left( \mathit{s}\right) }=0$ and $\mathit{d}_{\beta
}^{\left( \mathit{f}\right) }=1$ (half--filling). Doping the system with
additional fermions (holes in the case of cuprates) means that $\mathit{d}%
_{\beta }>0.5$. This increases the density $\mathit{d}_{\beta }^{\left(
\mathit{s}\right) }$ of the \textquotedblleft $\mathit{s}$\textquotedblright
--band which then becomes superconducting, whereas the \textquotedblleft $%
\mathit{f}$\textquotedblright --band stays half--filled, i.e., $\mathit{d}%
_{\beta }^{\left( \mathit{f}\right) }=1$. If the total density $\mathit{d}%
_{\beta }$ is too large ($\mathit{d}_{\beta }\gtrsim 0.73$) then
superconductivity is progressively suppressed and the \textquotedblleft $%
\mathit{f}$\textquotedblright --band is not anymore half--filled, i.e., $%
\mathit{d}_{\beta }^{\left( \mathit{f}\right) }>1$. The physical properties
of this (poorly superconducting) phase with $\mathit{d}_{\beta
}=1-\varepsilon $ ($\varepsilon =o(1)>0$) are not the same as in the
non--superconducting phase with $\mathit{d}_{\beta }=0.5$. The latter is a
purely Mott--insulating phase of \textquotedblleft $\mathit{f}$%
\textquotedblright --fermions and the first is a mixture (with fractions
depending on the given density $\mathit{d}_{\beta }$) \ of a purely
Mott--insulating phase of \textquotedblleft $\mathit{f}$\textquotedblright
--fermions with a phase describing a full \textquotedblleft $\mathit{f}$%
\textquotedblright --band. This is an interesting observation since the
excess of doping of holes in real cuprates leads to the destruction of
superconductivity and the appearance of a metallic phase, undoped cuprates
being Mott insulators. Of course, in our case it is not clear what the
thermodynamic phase with $\mathit{d}_{\beta }=1-\varepsilon $ has to do with
a metal since no kinetic energy is included and no transport properties are
analyzed.

Observe further that such a property is particular to the given choice of
parameters. See, e.g., Fig. \ref{mufixed1025.eps}, which corresponds to
choose $\delta =0.25$ and $\gamma _{\mathit{s}}=3.2$ (instead of $0.5$ and $%
3.7$) and where such a phenomenon is not observed around $\mathit{d}_{\beta
}\lessapprox 1$.
%TCIMACRO{%
%\TeXButton{mufixed1025.eps}{\begin{figure}[hbtp]
%\includegraphics[angle=0,scale=1,clip=true,width=6.5cm]{mufixed1025}
%\caption{\emph{Illustration
%of the Cooper pair condensate density $\textit{r}_{\beta }$ (red), the
%\textquotedblleft $\textit{s}$\textquotedblright--fermion\ density $\textit{d}_{\beta }^{\left( \textit{s}\right) }$ (orange) and the
%\textquotedblleft $\textit{f}$\textquotedblright--fermion\ density $\textit{d}_{\beta
%}^{\left( \textit{f}\right) }$ (blue) for $\beta=200$, $\gamma _{\textit{s}}=3.2$, $\delta =0.25$, $\lambda _{\textit{s}}=\lambda _{\textit{f}}=1$, $\lambda =0.85
%$, $h_{\textit{s}}=h_{\textit{f}}=\eta=0$, and $\mu _{\textit{s}}\in \left[ -2,7\right] $. The dashed green line corresponds
%to the average density $\textit{d}_{\beta }:=(\textit{d}_{\beta }^{\left( \textit{s}\right) }+\textit{d}_{\beta }^{\left( \textit{f}\right) })/2$. We concentrate our
%study between both vertical dashed black lines $\mu _{\textit{s}}=-1$ and $\mu _{\textit{s}}=2.5$.}}
%\label{mufixed1025.eps}
%\end{figure}}}%
%BeginExpansion
%EndExpansion

\begin{figure}[hbtp]
\includegraphics[angle=0,scale=1,clip=true,width=6.5cm]{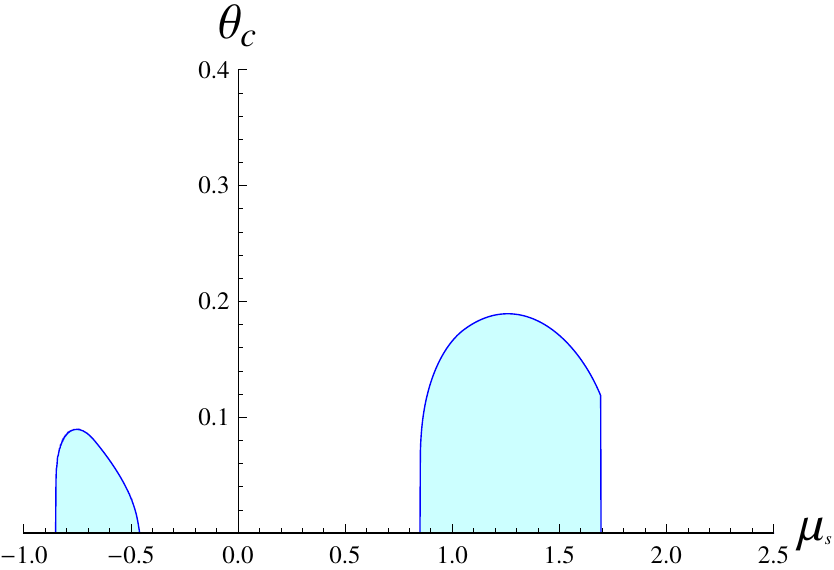}
\caption{\emph{Illustration
of the critical temperature $\theta _{c}$ for $\gamma _{\textit{s}}=3.7$, $\delta =0.5$, $\lambda _{\textit{s}}=\lambda _{\textit{f}}=1$, $\lambda =0.85
$, $h_{\textit{s}}=h_{\textit{f}}=\eta=0$, and $\mu _{\textit{s}}\in \left[ -1,2.5\right] $.
}}
\label{magncritical000d1.eps}
\end{figure}

Consider again the case $\delta =0.5$ and $\lambda _{\mathit{f}}=1$ of Fig. %
\ref{mufixed105.eps}. In contrast to a positive increase of $\mathit{d}%
_{\beta }$, one does not get a purely superconducting phase by decreasing
the total density $\mathit{d}_{\beta }$ of fermions. In fact, similarly to
\cite[Thm 6.5 (ii)]{BruPedra1} one sees from Fig. \ref{mufixed105.eps} that
we have always coexistence of Mott--insulating and superconducting phases
for $\mathit{d}_{\beta }\in \left[ 0.2,0.5\right] $ because of the first
order phase transition and density constraints (see below). The fraction of
the superconducting phase grows linearly from $0\%$ to $100\%$ when $\mathit{%
d}_{\beta }$ goes down from $0.5$ to approximately $0.2$. Moreover, as in
the other cases we find an asymmetry between superconductivity below and
above $\mathit{d}_{\beta }=0.5$ for the half--filled \textquotedblleft $%
\mathit{f}$\textquotedblright --band: It is easier to create a
superconducting phase by increasing $\mathit{d}_{\beta }$ than by decreasing
$\mathit{d}_{\beta }$, see Fig. \ref{magncritical000d1.eps}. As explained in
the case (a), this is due to the (screened) Coulomb interaction (\ref%
{coulomb repulsionbis}) in interplay with thermal excitations of
\textquotedblleft $\mathit{f}$\textquotedblright --fermions\ into the --
here energetically higher -- \textquotedblleft $\mathit{s}$%
\textquotedblright --level. The latter can be seen from Fig. \ref%
{mufixed105bisbis.eps} which corresponds to the inverse temperature $\beta
=11$.
%TCIMACRO{%
%\TeXButton{magncritical000d1.eps}{\begin{figure}[hbtp]
%\includegraphics[angle=0,scale=1,clip=true,width=6.5cm]{magncritical000d1}
%\caption{\emph{Illustration
%of the critical temperature $\theta _{c}$ for $\gamma _{\textit{s}}=3.7$, $\delta =0.5$, $\lambda _{\textit{s}}=\lambda _{\textit{f}}=1$, $\lambda =0.85
%$, $h_{\textit{s}}=h_{\textit{f}}=\eta=0$, and $\mu _{\textit{s}}\in \left[ -1,2.5\right] $.
%}}
%\label{magncritical000d1.eps}
%\end{figure}}}%
%BeginExpansion
%EndExpansion
%TCIMACRO{%
%\TeXButton{mufixed105bisbis.eps}{\begin{figure}[hbtp]
%\includegraphics[angle=0,scale=1,clip=true,width=6.5cm]{mufixed105bisbis}
%\caption{\emph{Illustration
%of the Cooper pair condensate density $\textit{r}_{\beta }$ (red), the
%\textquotedblleft $\textit{s}$\textquotedblright--fermion\ density $\textit{d}_{\beta }^{\left( \textit{s}\right) }$ (orange), and the
%\textquotedblleft $\textit{f}$\textquotedblright--fermion\ density $\textit{d}_{\beta
%}^{\left( \textit{f}\right) }$ (blue) for $\beta=11$, $\gamma _{\textit{s}}=3.7$, $\delta =0.5$, $\lambda _{\textit{s}}=\lambda _{\textit{f}}=1$, $\lambda =0.85
%$, $h_{\textit{s}}=h_{\textit{f}}=\eta=0$, and $\mu _{\textit{s}}\in \left[ -2,7\right] $. The dashed green line corresponds
%to the average density $\textit{d}_{\beta }:=(\textit{d}_{\beta }^{\left( \textit{s}\right) }+\textit{d}_{\beta }^{\left( \textit{f}\right) })/2$. We concentrate our
%study between both vertical dashed black lines $\mu _{\textit{s}}=-1$ and $\mu _{\textit{s}}=2.5$.}}
%\label{mufixed105bisbis.eps}
%\end{figure}}}%
%BeginExpansion
\begin{figure}[hbtp]
\includegraphics[angle=0,scale=1,clip=true,width=6.5cm]{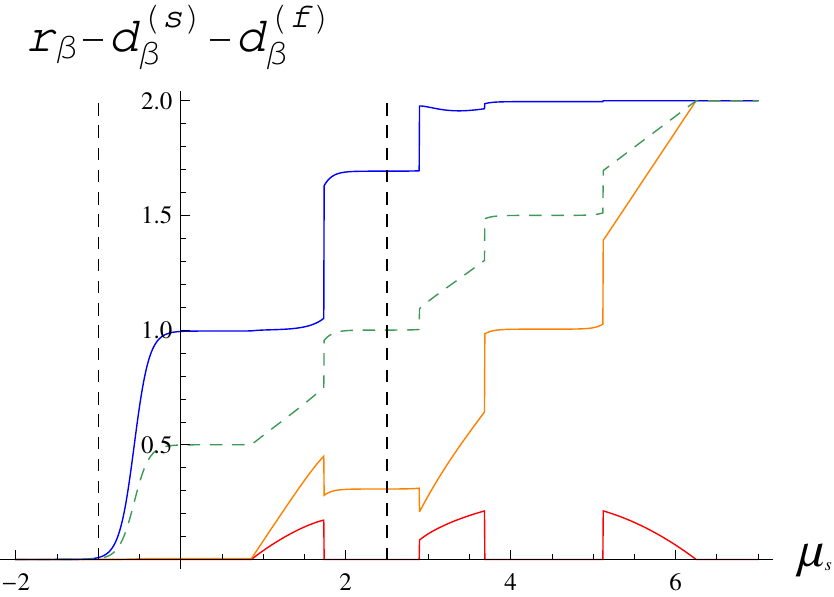}
\caption{\emph{Illustration
of the Cooper pair condensate density $\textit{r}_{\beta }$ (red), the
\textquotedblleft $\textit{s}$\textquotedblright--fermion\ density $\textit{d}_{\beta }^{\left( \textit{s}\right) }$ (orange), and the
\textquotedblleft $\textit{f}$\textquotedblright--fermion\ density $\textit{d}_{\beta
}^{\left( \textit{f}\right) }$ (blue) for $\beta=11$, $\gamma _{\textit{s}}=3.7$, $\delta =0.5$, $\lambda _{\textit{s}}=\lambda _{\textit{f}}=1$, $\lambda =0.85
$, $h_{\textit{s}}=h_{\textit{f}}=\eta=0$, and $\mu _{\textit{s}}\in \left[ -2,7\right] $. The dashed green line corresponds
to the average density $\textit{d}_{\beta }:=(\textit{d}_{\beta }^{\left( \textit{s}\right) }+\textit{d}_{\beta }^{\left( \textit{f}\right) })/2$. We concentrate our
study between both vertical dashed black lines $\mu _{\textit{s}}=-1$ and $\mu _{\textit{s}}=2.5$.}}
\label{mufixed105bisbis.eps}
\end{figure}%
%EndExpansion

By analyzing the model at fixed total densities $\rho \in (0,1)$ of fermions
(cf. (\ref{mu fixed particle density})) we obtain Fig. \ref%
{magncritical0d105bisD.eps} which represents the critical temperature as a
function of $\rho $. It is important to observe that Fig. \ref%
{magncritical0d105bisD.eps} is qualitatively in accordance with the
asymmetry experimentally found in cuprates. See, for instance, the schematic
phase diagram of real cuprates reproduced in \cite[Fig. 1]%
{Superconductivity3}.
%TCIMACRO{%
%\TeXButton{magncritical0d105bisD.eps}{\begin{figure}[hbtp]
%\includegraphics[angle=0,scale=1,clip=true,width=6.5cm]{magncritical0d105bisD}
%\caption{\emph{Illustration
%of the critical temperature $\theta _{c}$ for $\gamma _{\textit{s}}=3.7$, $\delta =0.5$, $\lambda _{\textit{s}}=\lambda _{\textit{f}}=1$, $\lambda =0.85
%$, $h_{\textit{s}}=h_{\textit{f}}=\eta=0$, and $\rho \in \left[ 0,1\right] $. The blue and yellow region correspond respectively
%to the superconducting and ferromagnetic--superconducting phases.
%The dashed line corresponds to the absence of doping.
%}}
%\label{magncritical0d105bisD.eps}
%\end{figure}}}%
%BeginExpansion
\begin{figure}[hbtp]
\includegraphics[angle=0,scale=1,clip=true,width=6.5cm]{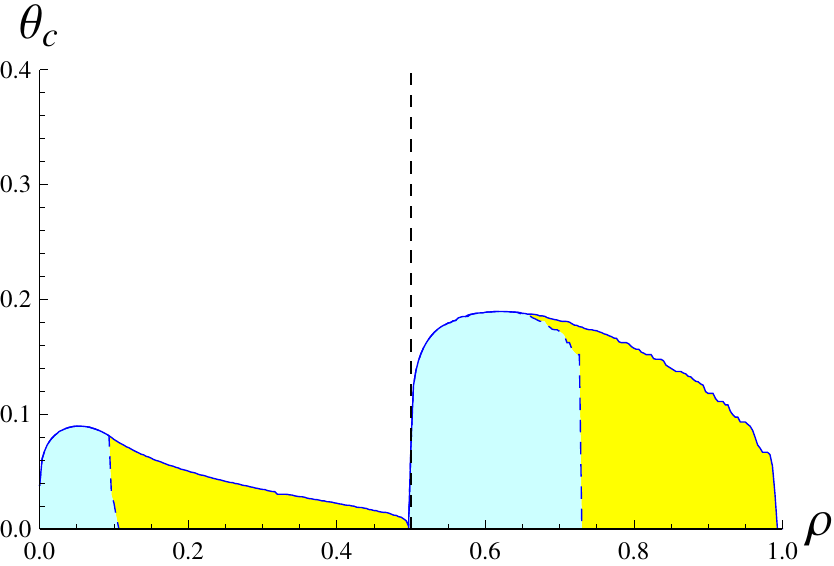}
\caption{\emph{Illustration
of the critical temperature $\theta _{c}$ for $\gamma _{\textit{s}}=3.7$, $\delta =0.5$, $\lambda _{\textit{s}}=\lambda _{\textit{f}}=1$, $\lambda =0.85
$, $h_{\textit{s}}=h_{\textit{f}}=\eta=0$, and $\rho \in \left[ 0,1\right] $. The blue and yellow region correspond respectively
to the superconducting and ferromagnetic--superconducting phases.
The dashed line corresponds to the absence of doping.
}}
\label{magncritical0d105bisD.eps}
\end{figure}%
%EndExpansion

\begin{figure}[hbtp]
\includegraphics[angle=0,scale=1,clip=true,width=6.5cm]{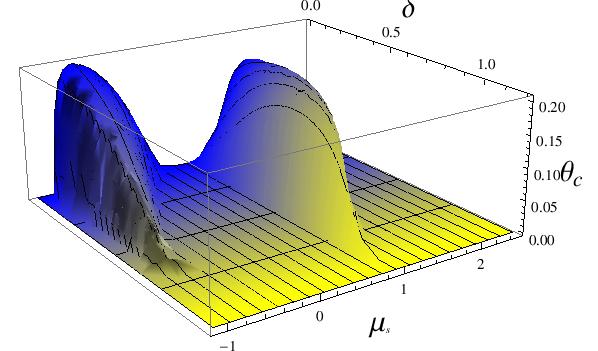}
\caption{\emph{Illustration
of the critical temperature $\theta _{c}$ for $\gamma _{\textit{s}}=3.7$,  $\lambda _{\textit{s}}=\lambda _{\textit{f}}=1$, $\lambda =0.85
$, $h_{\textit{s}}=h_{\textit{f}}=\eta=0$, $\mu _{\textit{s}}\in \left[ -1.1,2.5\right] $, and $\delta \in \left[ 0.,1.2\right] $. The color from blue to yellow reflects the increase of the
energy gap $|\delta|$.
}}
\label{plot3dgapbis2.eps}
\end{figure}

\begin{figure}[hbtp]
\includegraphics[angle=0,scale=1,clip=true,width=6.5cm]{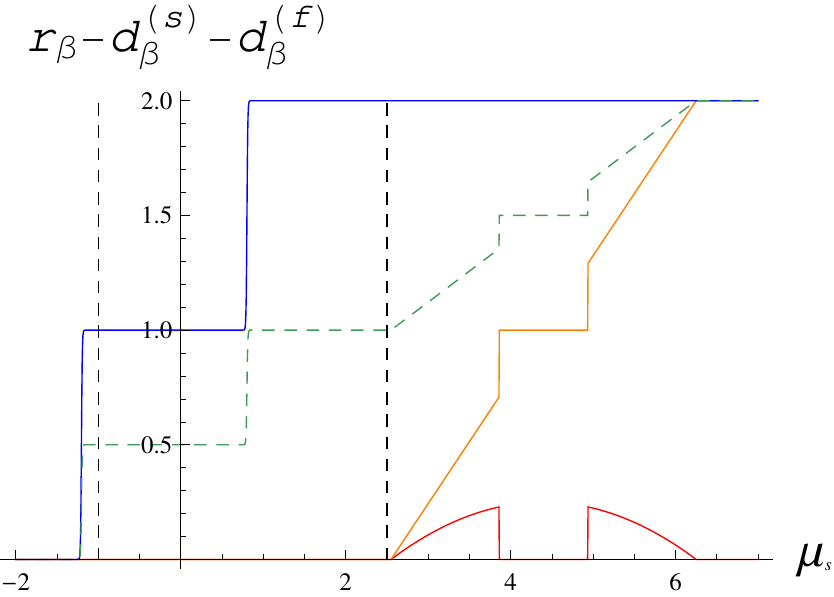}
\caption{\emph{Illustration
of the Cooper pair condensate density $\textit{r}_{\beta }$ (red), the
\textquotedblleft $\textit{s}$\textquotedblright--fermion\ density $\textit{d}_{\beta }^{\left( \textit{s}\right) }$ (orange), and the
\textquotedblleft $\textit{f}$\textquotedblright--fermion\ density $\textit{d}_{\beta
}^{\left( \textit{f}\right) }$ (blue) for $\beta=200$, $\gamma _{\textit{s}}=3.7$, $\delta =1.2$, $\lambda _{\textit{s}}=\lambda _{\textit{f}}=1$, $\lambda =0.85
$, $h_{\textit{s}}=h_{\textit{f}}=\eta=0$, and $\mu _{\textit{s}}\in \left[ -2,7\right] $. The dashed green line corresponds
to the average density $\textit{d}_{\beta }:=(\textit{d}_{\beta }^{\left( \textit{s}\right) }+\textit{d}_{\beta }^{\left( \textit{f}\right) })/2$. We concentrate our
study between both vertical dashed black lines $\mu _{\textit{s}}=-1$ and $\mu _{\textit{s}}=2.5$.}}
\label{mufixed112.eps}
\end{figure}

Finally, observe that other choices of parameters, specially of the energy
gap $\delta $, can completely modify the properties described here, see
Figs. \ref{plot3dgapbis2.eps}\ and \ref{mufixed112.eps}.\medskip\
%TCIMACRO{%
%\TeXButton{plot3dgapbis2.eps}{\begin{figure}[hbtp]
%\includegraphics[angle=0,scale=1,clip=true,width=6.5cm]{plot3dgapbis2}
%\caption{\emph{Illustration
%of the critical temperature $\theta _{c}$ for $\gamma _{\textit{s}}=3.7$,  $\lambda _{\textit{s}}=\lambda _{\textit{f}}=1$, $\lambda =0.85
%$, $h_{\textit{s}}=h_{\textit{f}}=\eta=0$, $\mu _{\textit{s}}\in \left[ -1.1,2.5\right] $, and $\delta \in \left[ 0.,1.2\right] $. The color from blue to yellow reflects the increase of the
%energy gap $|\delta|$.
%}}
%\label{plot3dgapbis2.eps}
%\end{figure}}}%
%BeginExpansion
%EndExpansion
%TCIMACRO{%
%\TeXButton{mufixed112.eps}{\begin{figure}[hbtp]
%\includegraphics[angle=0,scale=1,clip=true,width=6.5cm]{mufixed112}
%\caption{\emph{Illustration
%of the Cooper pair condensate density $\textit{r}_{\beta }$ (red), the
%\textquotedblleft $\textit{s}$\textquotedblright--fermion\ density $\textit{d}_{\beta }^{\left( \textit{s}\right) }$ (orange), and the
%\textquotedblleft $\textit{f}$\textquotedblright--fermion\ density $\textit{d}_{\beta
%}^{\left( \textit{f}\right) }$ (blue) for $\beta=200$, $\gamma _{\textit{s}}=3.7$, $\delta =1.2$, $\lambda _{\textit{s}}=\lambda _{\textit{f}}=1$, $\lambda =0.85
%$, $h_{\textit{s}}=h_{\textit{f}}=\eta=0$, and $\mu _{\textit{s}}\in \left[ -2,7\right] $. The dashed green line corresponds
%to the average density $\textit{d}_{\beta }:=(\textit{d}_{\beta }^{\left( \textit{s}\right) }+\textit{d}_{\beta }^{\left( \textit{f}\right) })/2$. We concentrate our
%study between both vertical dashed black lines $\mu _{\textit{s}}=-1$ and $\mu _{\textit{s}}=2.5$.}}
%\label{mufixed112.eps}
%\end{figure}}}%
%BeginExpansion
%EndExpansion

\textbf{(d) The almost full\ \textquotedblleft }$\mathit{f}$\textbf{%
\textquotedblright --band:}\medskip

\begin{figure}[hbtp]
\includegraphics[angle=0,scale=1,clip=true,width=6.5cm]{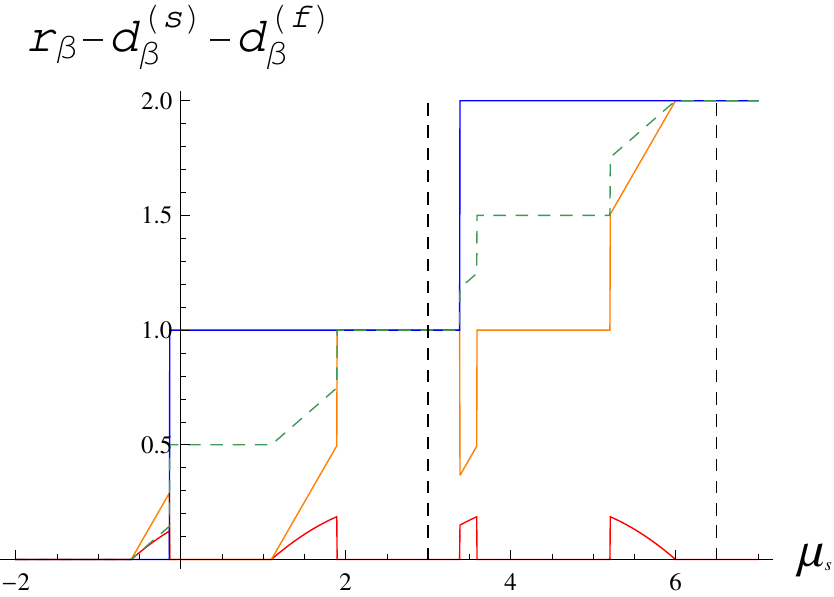}
\caption{\emph{Illustration
of the Cooper pair condensate density $\textit{r}_{\beta }$ (red), the
\textquotedblleft $\textit{s}$\textquotedblright--fermion\ density $\textit{d}_{\beta }^{\left( \textit{s}\right) }$ (orange), and the
\textquotedblleft $\textit{f}$\textquotedblright--fermion\ density $\textit{d}_{\beta
}^{\left( \textit{f}\right) }$ (blue) for $\beta=200$, $\gamma _{\textit{s}}=3.2$, $\delta =0.2$, $\lambda _{\textit{s}}=\lambda _{\textit{f}}=1$, $\lambda =0.85
$, $h_{\textit{s}}=h_{\textit{f}}=\eta=0$, and $\mu _{\textit{s}}\in \left[ -2,7\right] $. The dashed green line corresponds
to the average density $\textit{d}_{\beta }:=(\textit{d}_{\beta }^{\left( \textit{s}\right) }+\textit{d}_{\beta }^{\left( \textit{f}\right) })/2$. We concentrate our
study between both vertical dashed black lines $\mu _{\textit{s}}=3$ and $\mu _{\textit{s}}=6.5$.}}
\label{mufixed2.eps}
\end{figure}

This situation corresponds to Fig. \ref{mufixed2.eps}. The dashed black
lines correspond to the chemical potentials $\mu _{\mathit{s}}=3$ and $\mu _{%
\mathit{s}}=6.5$. In this case (d) we choose $\delta =0.2$, the other
parameters being the same as in cases (a) and (b). The electron--hole
asymmetry appears again and exactly for the same reasons, see Figs. \ref%
{magncritical0d2bis.eps} and \ref{mufixed2bis.eps}. Observe, however, that
the fermion density in the \textquotedblleft $\mathit{f}$\textquotedblright
\textbf{--}band is two (instead of zero or one) at low temperatures. Notice
also that -- in contrast to the case (a) -- the maximal critical
temperatures are always attained at $\mathit{d}_{\beta }^{\left( \mathit{s}%
\right) }>1$. At high enough temperatures the superconducting phase
completely disappears for $\mathit{d}_{\beta }^{\left( \mathit{s}\right) }<1$%
, whereas it can still be found if $\mathit{d}_{\beta }^{\left( \mathit{s}%
\right) }>1$. I.e., in this situation (d) superconductivity above
half--filling is more favored than superconductivity below half--filling. In
fact, if $\delta >0$ and the energetically lower level \textquotedblleft $%
\mathit{f}$\textquotedblright\ is full at zero temperature then \emph{no}
choice of parameters for which the maximum critical temperature is attained
at $\mathrm{d}_{\beta }^{\left( \mathit{s}\right) }<1$ were found.
%TCIMACRO{%
%\TeXButton{mufixed2.eps}{\begin{figure}[hbtp]
%\includegraphics[angle=0,scale=1,clip=true,width=6.5cm]{mufixed2}
%\caption{\emph{Illustration
%of the Cooper pair condensate density $\textit{r}_{\beta }$ (red), the
%\textquotedblleft $\textit{s}$\textquotedblright--fermion\ density $\textit{d}_{\beta }^{\left( \textit{s}\right) }$ (orange), and the
%\textquotedblleft $\textit{f}$\textquotedblright--fermion\ density $\textit{d}_{\beta
%}^{\left( \textit{f}\right) }$ (blue) for $\beta=200$, $\gamma _{\textit{s}}=3.2$, $\delta =0.2$, $\lambda _{\textit{s}}=\lambda _{\textit{f}}=1$, $\lambda =0.85
%$, $h_{\textit{s}}=h_{\textit{f}}=\eta=0$, and $\mu _{\textit{s}}\in \left[ -2,7\right] $. The dashed green line corresponds
%to the average density $\textit{d}_{\beta }:=(\textit{d}_{\beta }^{\left( \textit{s}\right) }+\textit{d}_{\beta }^{\left( \textit{f}\right) })/2$. We concentrate our
%study between both vertical dashed black lines $\mu _{\textit{s}}=3$ and $\mu _{\textit{s}}=6.5$.}}
%\label{mufixed2.eps}
%\end{figure}}}%
%BeginExpansion
%EndExpansion
%TCIMACRO{%
%\TeXButton{magncritical0d2bis.eps}{\begin{figure}[hbtp]
%\includegraphics[angle=0,scale=1,clip=true,width=6.5cm]{magncritical0d2bis}
%\caption{\emph{Illustration
%of the critical temperature $\theta _{c}$ for $\gamma _{\textit{s}}=3.2$, $\delta =0.2$, $\lambda _{\textit{s}}=\lambda _{\textit{f}}=1$, $\lambda =0.85
%$, $h_{\textit{s}}=h_{\textit{f}}=\eta=0$, and $\mu _{\textit{s}}\in \left[ 3,6.5\right] $.
%}}
%\label{magncritical0d2bis.eps}
%\end{figure}}}%
%BeginExpansion
\begin{figure}[hbtp]
\includegraphics[angle=0,scale=1,clip=true,width=6.5cm]{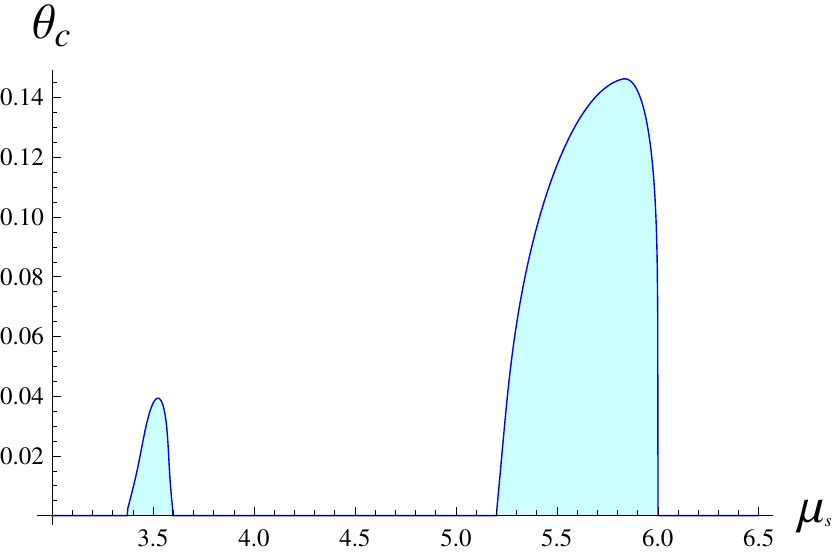}
\caption{\emph{Illustration
of the critical temperature $\theta _{c}$ for $\gamma _{\textit{s}}=3.2$, $\delta =0.2$, $\lambda _{\textit{s}}=\lambda _{\textit{f}}=1$, $\lambda =0.85
$, $h_{\textit{s}}=h_{\textit{f}}=\eta=0$, and $\mu _{\textit{s}}\in \left[ 3,6.5\right] $.
}}
\label{magncritical0d2bis.eps}
\end{figure}%
%EndExpansion
%TCIMACRO{%
%\TeXButton{mufixed2bis.eps}{\begin{figure}[hbtp]
%\includegraphics[angle=0,scale=1,clip=true,width=6.5cm]{mufixed2bis}
%\caption{\emph{Illustration
%of the Cooper pair condensate density $\textit{r}_{\beta }$ (red), the
%\textquotedblleft $\textit{s}$\textquotedblright--fermion\ density $\textit{d}_{\beta }^{\left( \textit{s}\right) }$ (orange), and the
%\textquotedblleft $\textit{f}$\textquotedblright--fermion\ density $\textit{d}_{\beta
%}^{\left( \textit{f}\right) }$ (blue) for $\beta=28$, $\gamma _{\textit{s}}=3.2$, $\delta =0.2$, $\lambda _{\textit{s}}=\lambda _{\textit{f}}=1$, $\lambda =0.85
%$, $h_{\textit{s}}=h_{\textit{f}}=\eta=0$, and $\mu _{\textit{s}}\in \left[ -2,7\right] $. The dashed green line corresponds
%to the average density $\textit{d}_{\beta }:=(\textit{d}_{\beta }^{\left( \textit{s}\right) }+\textit{d}_{\beta }^{\left( \textit{f}\right) })/2$. We concentrate our
%study between both vertical dashed black lines $\mu _{\textit{s}}=3$ and $\mu _{\textit{s}}=6.5$.}}
%\label{mufixed2bis.eps}
%\end{figure}}}%
%BeginExpansion
\begin{figure}[hbtp]
\includegraphics[angle=0,scale=1,clip=true,width=6.5cm]{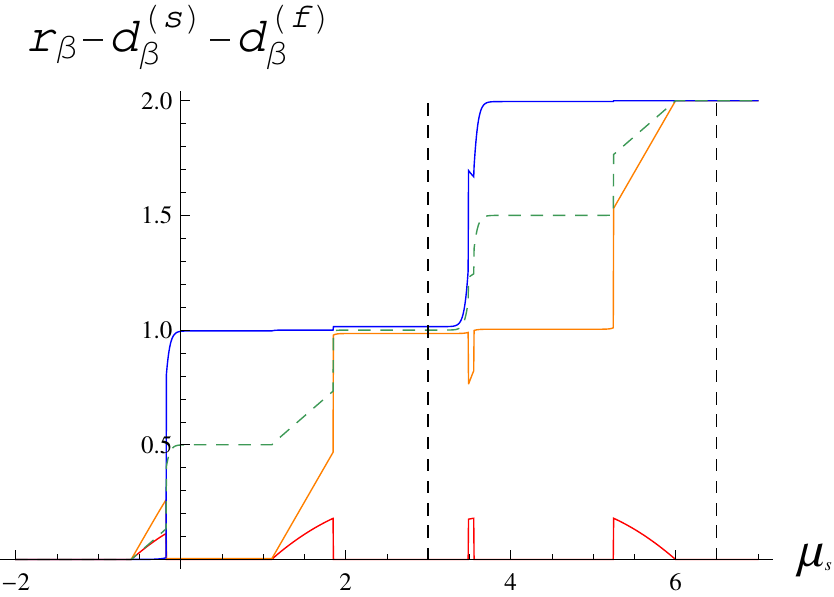}
\caption{\emph{Illustration
of the Cooper pair condensate density $\textit{r}_{\beta }$ (red), the
\textquotedblleft $\textit{s}$\textquotedblright--fermion\ density $\textit{d}_{\beta }^{\left( \textit{s}\right) }$ (orange), and the
\textquotedblleft $\textit{f}$\textquotedblright--fermion\ density $\textit{d}_{\beta
}^{\left( \textit{f}\right) }$ (blue) for $\beta=28$, $\gamma _{\textit{s}}=3.2$, $\delta =0.2$, $\lambda _{\textit{s}}=\lambda _{\textit{f}}=1$, $\lambda =0.85
$, $h_{\textit{s}}=h_{\textit{f}}=\eta=0$, and $\mu _{\textit{s}}\in \left[ -2,7\right] $. The dashed green line corresponds
to the average density $\textit{d}_{\beta }:=(\textit{d}_{\beta }^{\left( \textit{s}\right) }+\textit{d}_{\beta }^{\left( \textit{f}\right) })/2$. We concentrate our
study between both vertical dashed black lines $\mu _{\textit{s}}=3$ and $\mu _{\textit{s}}=6.5$.}}
\label{mufixed2bis.eps}
\end{figure}%
%EndExpansion
\medskip

Cases (b) and (d) can also be considered -- as it was done for the cases (a)
and (c) -- w.r.t. a fixed total fermion density $\rho \in (0,2)$ (instead of
a fixed chemical potential $\mu _{\mathit{s}}$), see (\ref{mu fixed particle
density}). All phenomena concerning the electron--hole asymmetry described
at fixed $\mu _{\mathit{s}}$ also appear at fixed $\rho \in (0,2)$
completely analogously to the cases (a) and (c). A detailed analysis is
therefore omitted.

From the physical description given at the beginning of Section \ref{section
II}, cuprates seem to fit in case (c) since $\delta >0$ whereas $\mathit{d}%
_{\beta }^{\left( \mathit{f}\right) }=1$ and $\mathit{d}_{\beta }^{\left(
\mathit{s}\right) }=0$ at $\mathit{d}_{\beta }=0.5$ (no doping). In this
case the non--superconducting band (\textquotedblleft $\mathit{f}$%
\textquotedblright \textbf{--}band) should be identified with copper
orbitals -- which is half--filled without doping -- and the superconducting
band (\textquotedblleft $\mathit{s}$\textquotedblright \textbf{--}band) with
oxygen orbitals -- which is empty without doping. Note that experiments
indicate that superconduction takes place in oxygen orbitals (see Section
5.11 of \cite{Saxena}). Considering Hamiltonians for holes in cuprates, our
model correctly predicts (see case (c)) that if $\delta >0$ (i.e., if holes
in oxygen orbitals are higher in energy than holes in copper orbitals, which
is the case in real cuprates), then the highest critical temperature is
obtained at $\mathit{d}_{\beta }^{\left( \mathit{s}\right) }>1$, i.e., by
introducing donors of holes in the cuprate crystal.

\begin{figure}[hbtp]
\includegraphics[angle=0,scale=1,clip=true,width=6.5cm]{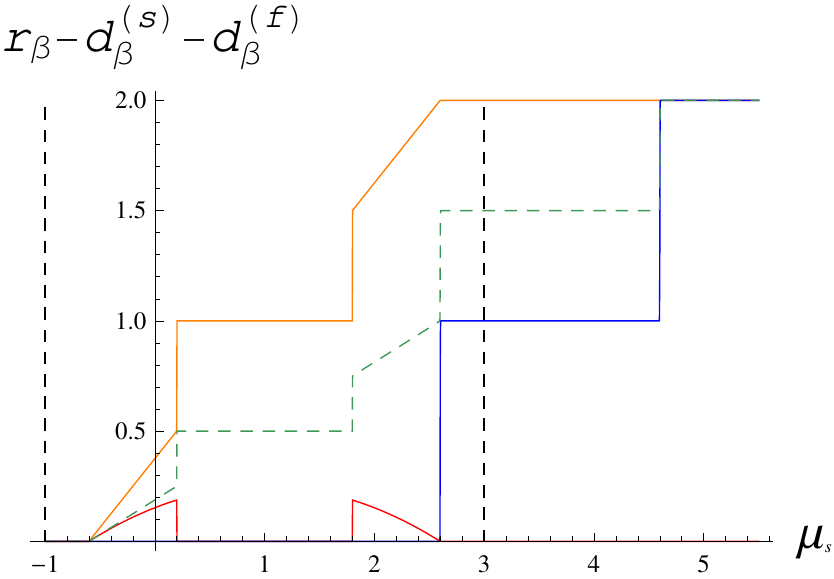}
\caption{\emph{Illustration
of the Cooper pair condensate density $\textit{r}_{\beta }$ (red), the
\textquotedblleft $\textit{s}$\textquotedblright--fermion\ density
$\textit{d}_{\beta }^{\left( \textit{s}\right) }$ (orange), and the
\textquotedblleft $\textit{f}$\textquotedblright--fermion\ density $\textit{d}_{\beta
}^{\left( \textit{f}\right) }$ (blue) for $\beta=2000$, $\gamma _{\textit{s}}=3.2$, $\delta =-2.6$, $\lambda _{\textit{s}}=\lambda _{\textit{f}}=1$,
$\eta=0.5$, $h_{\textit{s}}=h_{\textit{f}}=\lambda=0$, and $\mu _{\textit{s}}\in \left[ -1,5.5\right] $. The dashed green line corresponds
to the average density $\textit{d}_{\beta }:=(\textit{d}_{\beta }^{\left( \textit{s}\right) }+\textit{d}_{\beta }^{\left( \textit{f}\right) })/2$. We concentrate our
study between both vertical dashed black lines $\mu _{\textit{s}}=-1$ and $\mu _{\textit{s}}=3$.}}
\label{mufixeL0.eps}
\end{figure}

Note finally that the phenomenon of electron--hole asymmetry can also take
place if the inter--band interaction is purely magnetic (instead of purely
electrostatic) or is mixed (i.e., has magnetic and electrostatic
components). As in cases (a) or (d) one can find examples where the maximal
critical temperature of superconductivity is either at $\mathit{d}_{\beta
}^{\left( \mathit{s}\right) }<1$ or at $\mathit{d}_{\beta }^{\left( \mathit{s%
}\right) }>1$. See, e.g., Figs. \ref{mufixeL0.eps} and \ref%
{magncritical0LO.eps}. Moreover, in the case of a non--vanishing magnetic
inter--band interaction%
\begin{equation}
-\eta \sum_{x\in \Lambda _{N}}(n_{x,\uparrow }^{(\mathit{s}%
)}-n_{x,\downarrow }^{(\mathit{s})})(n_{x,\uparrow }^{(\mathit{f}%
)}-n_{x,\downarrow }^{(\mathit{f})})  \label{magnetic interaction}
\end{equation}%
a new phenomenon takes place:\ The so--called reentering behavior which is
analyzed below in more details.
%TCIMACRO{%
%\TeXButton{mufixeL0.eps}{\begin{figure}[hbtp]
%\includegraphics[angle=0,scale=1,clip=true,width=6.5cm]{mufixeL0}
%\caption{\emph{Illustration
%of the Cooper pair condensate density $\textit{r}_{\beta }$ (red), the
%\textquotedblleft $\textit{s}$\textquotedblright--fermion\ density
%$\textit{d}_{\beta }^{\left( \textit{s}\right) }$ (orange), and the
%\textquotedblleft $\textit{f}$\textquotedblright--fermion\ density $\textit{d}_{\beta
%}^{\left( \textit{f}\right) }$ (blue) for $\beta=2000$, $\gamma _{\textit{s}}=3.2$, $\delta =-2.6$, $\lambda _{\textit{s}}=\lambda _{\textit{f}}=1$,
%$\eta=0.5$, $h_{\textit{s}}=h_{\textit{f}}=\lambda=0$, and $\mu _{\textit{s}}\in \left[ -1,5.5\right] $. The dashed green line corresponds
%to the average density $\textit{d}_{\beta }:=(\textit{d}_{\beta }^{\left( \textit{s}\right) }+\textit{d}_{\beta }^{\left( \textit{f}\right) })/2$. We concentrate our
%study between both vertical dashed black lines $\mu _{\textit{s}}=-1$ and $\mu _{\textit{s}}=3$.}}
%\label{mufixeL0.eps}
%\end{figure}}}%
%BeginExpansion
%EndExpansion
%TCIMACRO{%
%\TeXButton{magncritical0LO.eps}{\begin{figure}[hbtp]
%\includegraphics[angle=0,scale=1,clip=true,width=6.5cm]{magncritical0LO}
%\caption{\emph{Illustration
%of the critical temperature $\theta _{c}$ for $\gamma _{\textit{s}}=3.2$, $\delta =-2.6$, $\lambda _{\textit{s}}=\lambda _{\textit{f}}=1$,
%$\eta=0.5$, $h_{\textit{s}}=h_{\textit{f}}=\lambda=0$, and $\mu _{\textit{s}}\in \left[ -1,3\right] $.
%}}
%\label{magncritical0LO.eps}
%\end{figure}}}%
%BeginExpansion
\begin{figure}[hbtp]
\includegraphics[angle=0,scale=1,clip=true,width=6.5cm]{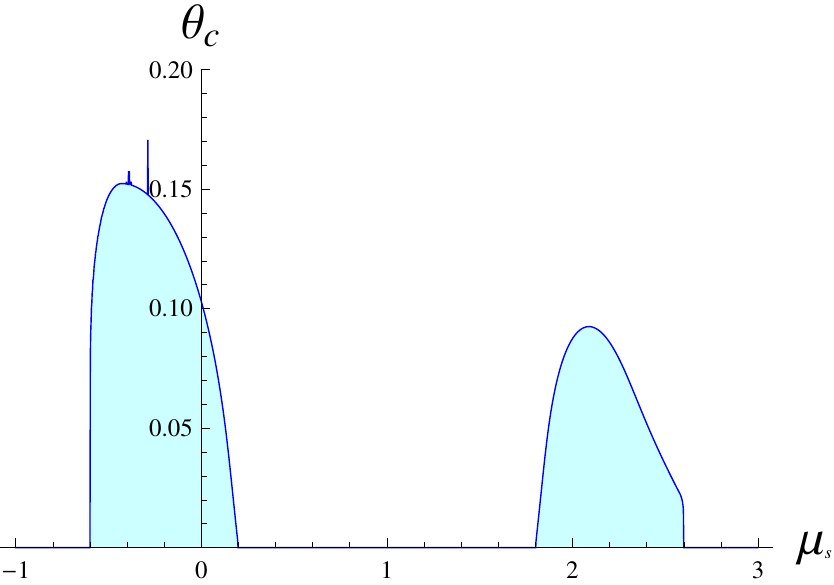}
\caption{\emph{Illustration
of the critical temperature $\theta _{c}$ for $\gamma _{\textit{s}}=3.2$, $\delta =-2.6$, $\lambda _{\textit{s}}=\lambda _{\textit{f}}=1$,
$\eta=0.5$, $h_{\textit{s}}=h_{\textit{f}}=\lambda=0$, and $\mu _{\textit{s}}\in \left[ -1,3\right] $.
}}
\label{magncritical0LO.eps}
\end{figure}%
%EndExpansion

\section{Reentering behavior and magnetic inter--band interaction}

We fix $\eta \neq 0$ in (\ref{hamiltoninteraction}). Otherwise the
reentering behavior would not take place. Since we are only interested on
effects due to the magnetic inter--band interaction (\ref{magnetic
interaction}) on the superconductivity of \textquotedblleft $\mathit{s}$%
\textquotedblright \textbf{--}fermions we let $\lambda =0$ in (\ref%
{hamiltoninteraction}). A small electrostatic component $\lambda \simeq 0$
does not change the qualitative thermodynamic behavior of the model.

\begin{figure}[hbtp]
\includegraphics[angle=0,scale=1,clip=true,width=6.5cm]{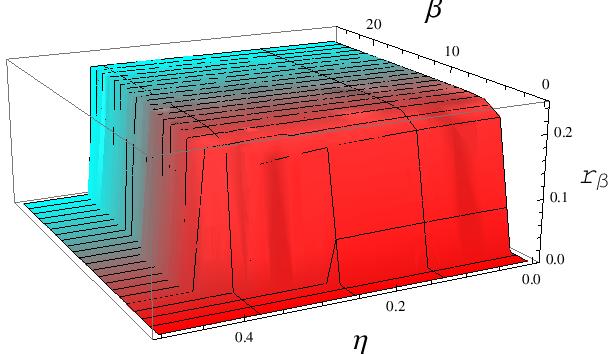}
\caption{\emph{Illustration, as a function of $\eta\in \left[ 0,0.5\right] $ and $\beta \in \left[ 0.1,25\right] $, of the Cooper pair condensate density $\textit{r}_{\beta }$ for $\gamma _{\textit{s}}=3.2$, $\lambda _{\textit{s}}=\lambda _{\textit{f}}=0.5$, $\mu _{\textit{s}}=\mu _{\textit{f}}=1$ and $h_{\textit{s}}=h_{\textit{f}}=\lambda =0$. The color from red to blue reflects the decrease of the
temperature. }}
\label{pferro1.eps}
\end{figure}

An illustration of the Cooper pair condensate density as well as the
critical temperature w.r.t. the chemical potential $\mu _{\mathit{s}}$ of
\textquotedblleft $\mathit{s}$\textquotedblright --fermions \ is given for
this situation in Figs. \ref{pferro1.eps} and \ref%
{magncriticalgraph1bism.eps}. It is interesting to observe that a reentering
behavior occurs: At $\mu _{\mathit{s}}=0.4$ (dashed line in Fig. \ref%
{magncriticalgraph1bism.eps}) the system becomes superconducting below a
critical temperature $\theta _{c_{1}}$ and then leaves the superconducting
phase below $\theta _{c_{2}}$ with $\theta _{c_{2}}<\theta _{c_{1}}$. This
kind of phenomenology is found in real ferromagnetic superconductors \cite[%
p. 263-267]{Shrivastava}.
%TCIMACRO{%
%\TeXButton{pferro1.eps}{\begin{figure}[hbtp]
%\includegraphics[angle=0,scale=1,clip=true,width=6.5cm]{pferro1}
%\caption{\emph{Illustration, as a function of $\eta\in \left[ 0,0.5\right] $ and $\beta \in \left[ 0.1,25\right] $, of the Cooper pair condensate density $\textit{r}_{\beta }$ for $\gamma _{\textit{s}}=3.2$, $\lambda _{\textit{s}}=\lambda _{\textit{f}}=0.5$, $\mu _{\textit{s}}=\mu _{\textit{f}}=1$ and $h_{\textit{s}}=h_{\textit{f}}=\lambda =0$. The color from red to blue reflects the decrease of the
%temperature. }}
%\label{pferro1.eps}
%\end{figure}}}%
%BeginExpansion
%EndExpansion

%TCIMACRO{%
%\TeXButton{magncriticalgraph1bism.eps}{\begin{figure}[hbtp]
%\includegraphics[angle=0,scale=1,clip=true,width=6.5cm]{magncriticalgraph1bism}
%\caption{\emph{Illustration of the domain of
%temperature (blue region) with a superconducting phase for $\gamma _{\textit{s}}=3.2$, $\lambda _{\textit{s}}=\lambda _{\textit{f}}=0.5$, $\mu _{\textit{s}}=\mu _{\textit{f}}=1$, $h_{\textit{s}}=h_{\textit{f}}=\lambda =0$, and $\eta\in \left[ 0,0.5\right] $. For, e.g., $\eta=0.4$ (black dashed line)
%a reentering behaviour appears.}}
%\label{magncriticalgraph1bism.eps}
%\end{figure}}}%
%BeginExpansion
\begin{figure}[hbtp]
\includegraphics[angle=0,scale=1,clip=true,width=6.5cm]{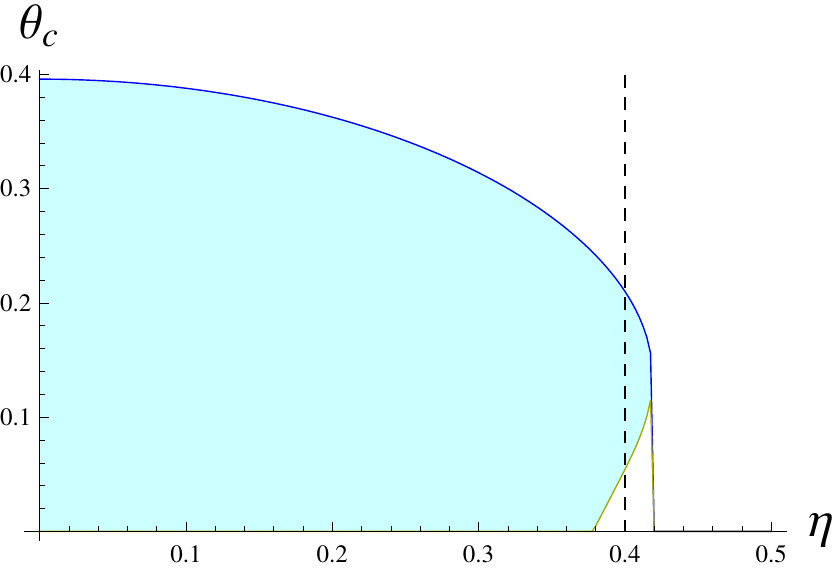}
\caption{\emph{Illustration of the domain of
temperature (blue region) with a superconducting phase for $\gamma _{\textit{s}}=3.2$, $\lambda _{\textit{s}}=\lambda _{\textit{f}}=0.5$, $\mu _{\textit{s}}=\mu _{\textit{f}}=1$, $h_{\textit{s}}=h_{\textit{f}}=\lambda =0$, and $\eta\in \left[ 0,0.5\right] $. For, e.g., $\eta=0.4$ (black dashed line)
a reentering behaviour appears.}}
\label{magncriticalgraph1bism.eps}
\end{figure}%
%EndExpansion

\begin{figure}[hbtp]
\includegraphics[angle=0,scale=1,clip=true,width=6.5cm]{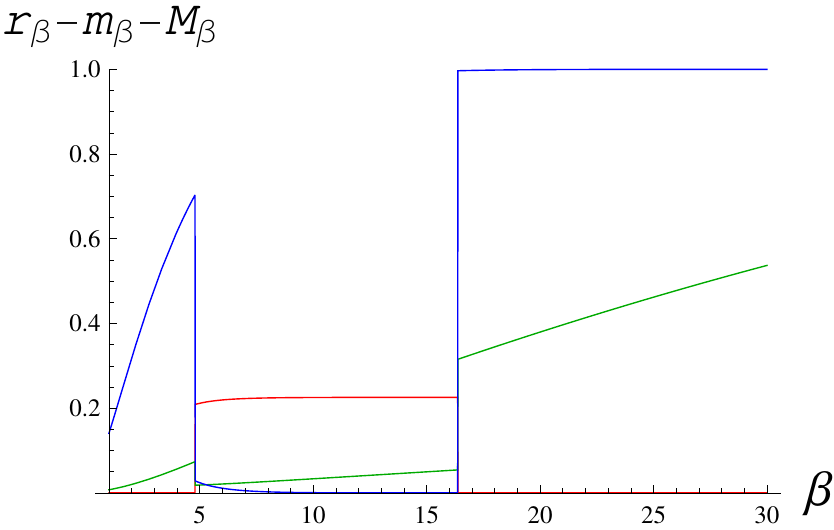}
\caption{\emph{Illustration of the Cooper pair condensate density $\textit{r}_{\beta }$ (red),
the intra-band magnetization density $\mathrm{m}_{\beta }$ (green),
and the inter-band magnetization density $\mathrm{M}_{\beta }$ (blue)
for $\gamma _{\textit{s}}=3.2$, $\lambda _{\textit{s}}=\lambda _{\textit{f}}=0.5$, $\mu _{\textit{s}}=\mu _{\textit{f}}=1$, $\eta=0.4$, $h_{\textit{s}}=h_{\textit{f}}=0.01$, $\lambda =0$, and $\beta\in \left[ 1,30\right] $. }}
\label{pferro3.eps}
\end{figure}

The total magnetization density per band%
\begin{equation*}
\mathit{m}_{\beta }:=\underset{N\rightarrow \infty }{\lim }\omega _{N}\left(
n_{x,\uparrow }^{(\mathit{s})}-n_{x,\downarrow }^{(\mathit{s}%
)}+n_{x,\uparrow }^{(\mathit{f})}-n_{x,\downarrow }^{(\mathit{f})}\right)
\end{equation*}%
($\mathit{m}_{\beta }\in \lbrack -1,1]$) equals%
\begin{eqnarray}
\mathit{m}_{\beta } &=&\frac{e^{-2\beta \lambda }}{\mathrm{f}(\mathit{r}%
_{\beta })}\Bigl\{4\mathrm{e}^{\beta \eta }\sinh \left( \beta \left( h_{%
\mathit{f}}+h_{\mathit{s}}\right) \right)  \notag \\
&&+(\mathrm{e}^{-\beta \left( \mu _{\mathit{f}}-2\lambda \right) }+\mathrm{e}%
^{-\beta (2\lambda +2\lambda _{\mathit{f}}-\mu _{\mathit{f}})})\sinh (\beta
h_{\mathit{s}})  \notag \\
&&+2\mathrm{e}^{-\beta \lambda _{\mathit{s}}}\cosh (\beta g_{r,2\lambda
})\sinh (\beta h_{\mathit{f}})\Bigr\}  \label{eq 20}
\end{eqnarray}%
away from any critical point. In particular, -- in contrast to what is
observed in real ferromagnetic superconductors -- no spontaneous
magnetization can occur in this model, as the Hamiltonian is invariant under
exchange of $\uparrow $ and $\downarrow $ spins for $h_{\mathit{s}}=h_{%
\mathit{f}}=0$. A more complicated model including, for instance,
Heisenberg--type intra--band interaction terms could show spontaneous
magnetization. In order to simplify the analysis we induce magnetization by
imposing a small external magnetic field $h_{\mathit{f}}\neq 0.$ See Fig. %
\ref{pferro3.eps}. In this case we obtain some (poorly) ferromagnetic
superconducting phase at fixed chemical potential. In contrast to models
showing spontaneous magnetization and real ferromagnetic superconductors,
this procedure has of course the drawback of implying -- at any fixed
temperature -- small magnetizations at small fields $h_{\mathit{f}},h_{%
\mathit{s}}$, i.e., $|\mathit{m}_{\beta }|=O(|h_{\mathit{f}}|+|h_{\mathit{s}%
}|)$, see Fig. \ref{pferro3.eps}. For more details we recommend \cite[Sect.
3.3]{BruPedra1}.

Proceeding in this way the total magnetization density $\mathit{m}_{\beta }$
in the superconducting phase generally stays rather small: In Fig. \ref%
{pferro3.eps} $\mathit{m}_{\beta }$ is at most 23\% of $\mathit{r}_{\beta }$
in the superconducting phase. However, $\mathit{m}_{\beta }$ is increased by
almost 600\% at the lowest critical temperature of superconductivity. This
captures in a sense the property of real ferromagnetic superconductors of
going into a ferromagnetic phase when they leave the superconducting phase
at low temperatures (i.e., below the critical temperature $\theta _{c_{2}}$).

Note that, if the range of temperatures with a superconducting phase is
sufficiently small then the magnetization $\mathit{m}_{\beta }$ (in the
superconducting phase) becomes almost zero, see, e.g., \cite[Corollary 3.3]%
{BruPedra1}. There is also a critical magnetic field $|h_{%
\mathit{s}}^{\left( c\right) }|$ above which the superconducting phase
cannot exist at any temperature: This can be seen as (part of) the usual Mei{%
\ss }ner effect.
%TCIMACRO{%
%\TeXButton{pferro3.eps}{\begin{figure}[hbtp]
%\includegraphics[angle=0,scale=1,clip=true,width=6.5cm]{pferro3}
%\caption{\emph{Illustration of the Cooper pair condensate density $\textit{r}_{\beta }$ (red),
%the intra-band magnetization density $\mathrm{m}_{\beta }$ (green),
%and the inter-band magnetization density $\mathrm{M}_{\beta }$ (blue)
%for $\gamma _{\textit{s}}=3.2$, $\lambda _{\textit{s}}=\lambda _{\textit{f}}=0.5$, $\mu _{\textit{s}}=\mu _{\textit{f}}=1$, $\eta=0.4$, $h_{\textit{s}}=h_{\textit{f}}=0.01$, $\lambda =0$, and $\beta\in \left[ 1,30\right] $. }}
%\label{pferro3.eps}
%\end{figure}}}%
%BeginExpansion
%EndExpansion

\begin{figure}[hbtp]
\includegraphics[angle=0,scale=1,clip=true,width=6.5cm]{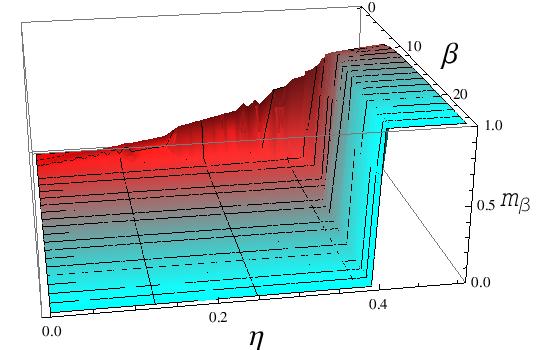}
\caption{\emph{Illustration of the inter--band magnetization density $\mathrm{M}_{\beta }$ for $\gamma _{\textit{s}}=3.2$, $\lambda _{\textit{s}}=\lambda _{\textit{f}}=0.5$, $\mu _{\textit{s}}=\mu _{\textit{f}}=1$ and $h_{\textit{s}}=h_{\textit{f}}=\lambda =0$, $\eta\in \left[ 0,0.5\right] $, and $\beta \in \left[ 0.1,25\right] $.
The color from red to blue reflects the decrease of the temperature. }}
\label{pferro1bis.eps}
\end{figure}

\begin{figure}[hbtp]
\includegraphics[angle=0,scale=1,clip=true,width=6.5cm]{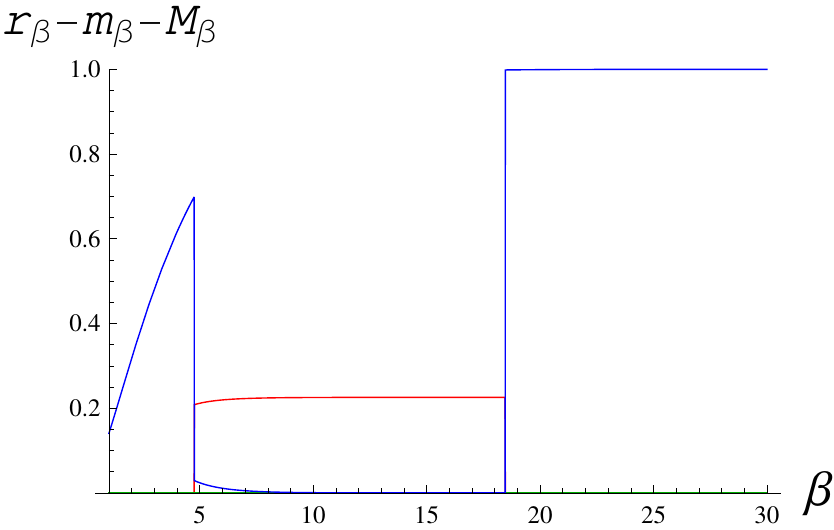}
\caption{\emph{Illustration of the Cooper pair condensate density $\textit{r}_{\beta }$ (red), the intra-band magnetization density $\mathrm{m}_{\beta }$ (green), and the inter-band magnetization density $\mathrm{M}_{\beta }$ (blue) for $\gamma _{\textit{s}}=3.2$, $\lambda _{\textit{s}}=\lambda _{\textit{f}}=0.5$, $\mu _{\textit{s}}=\mu _{\textit{f}}=1$, $\eta=0.4$, $h_{\textit{s}}=h_{\textit{f}}=\lambda =0$, and $\beta\in \left[ 1,30\right] $. }}
\label{pferro2.eps}
\end{figure}

The inter--band magnetization density
\begin{equation*}
\mathit{M}_{\beta }:=\underset{N\rightarrow \infty }{\lim }\omega _{N}\left(
(n_{x,\uparrow }^{(\mathit{f})}-n_{x,\downarrow }^{(\mathit{f}%
)})(n_{x,\uparrow }^{(\mathit{s})}-n_{x,\downarrow }^{(\mathit{s})})\right)
\end{equation*}%
($\mathit{M}_{\beta }\in \lbrack -1,1]$) equals%
\begin{eqnarray}
\mathit{M}_{\beta } &=&\frac{e^{-2\beta \lambda }}{\mathrm{f}(\mathit{r}%
_{\beta })}\Bigl\{\mathrm{e}^{-\beta h_{\mathit{s}}}\sinh (\beta (\eta -h_{%
\mathit{f}}))  \notag \\
&&+\mathrm{e}^{\beta h_{\mathit{s}}}\sinh \left( \beta (\eta +h_{\mathit{f}%
})\right) \Bigr\}  \label{griffiths5}
\end{eqnarray}%
away from any critical point. In the superconducting phase, 100\% of
\textquotedblleft $\mathit{s}$\textquotedblright --fermions\ form Cooper
pairs in the limit of zero--temperature ($\beta \rightarrow \infty $). This
implies that $\mathit{M}_{\beta }=0$ in this limit. See Figs. \ref%
{pferro1.eps} and \ref{pferro1bis.eps} as well as Fig. \ref{pferro2.eps}.
Moreover, because of symmetries of the model, the magnetic inter--band
interaction (\ref{magnetic interaction}) is always zero at $h_{\mathit{s}%
}=h_{\mathit{f}}=0$. Thus, one could think that the magnetic interaction
does not perturb much the superconducting phase of \textquotedblleft $%
\mathit{s}$\textquotedblright --fermions. It turns out that the
thermodynamic effect of the magnetic inter--band coupling is by far not
negligible: It can even completely destroy the superconducting phase, either
at low enough temperature, or for any temperature when $|\eta |>\eta
_{\infty }^{(c)}$. In particular, the two--band model discussed here shows a
reentering behavior also without any external magnetic field (and thus,
without ferromagnetism), see Fig. \ref{pferro2.eps}. Indeed, at any fixed
inverse temperature $\beta >0$ there is a critical magnetic inter--band
interaction $\eta _{\beta }^{(c)}$ such that the superconducting phase
disappears for $|\eta |>\eta _{\beta }^{(c)}$, even if there is no
magnetization, see Fig. \ref{magncriticalgraph1bism.eps}.
%TCIMACRO{%
%\TeXButton{pferro1bis.eps}{\begin{figure}[hbtp]
%\includegraphics[angle=0,scale=1,clip=true,width=6.5cm]{pferro1bis}
%\caption{\emph{Illustration of the inter--band magnetization density $\mathrm{M}_{\beta }$ for $\gamma _{\textit{s}}=3.2$, $\lambda _{\textit{s}}=\lambda _{\textit{f}}=0.5$, $\mu _{\textit{s}}=\mu _{\textit{f}}=1$ and $h_{\textit{s}}=h_{\textit{f}}=\lambda =0$, $\eta\in \left[ 0,0.5\right] $, and $\beta \in \left[ 0.1,25\right] $.
%The color from red to blue reflects the decrease of the temperature. }}
%\label{pferro1bis.eps}
%\end{figure}}}%
%BeginExpansion
%EndExpansion
%TCIMACRO{%
%\TeXButton{pferro2.eps}{\begin{figure}[hbtp]
%\includegraphics[angle=0,scale=1,clip=true,width=6.5cm]{pferro2}
%\caption{\emph{Illustration of the Cooper pair condensate density $\textit{r}_{\beta }$ (red), the intra-band magnetization density $\mathrm{m}_{\beta }$ (green), and the inter-band magnetization density $\mathrm{M}_{\beta }$ (blue) for $\gamma _{\textit{s}}=3.2$, $\lambda _{\textit{s}}=\lambda _{\textit{f}}=0.5$, $\mu _{\textit{s}}=\mu _{\textit{f}}=1$, $\eta=0.4$, $h_{\textit{s}}=h_{\textit{f}}=\lambda =0$, and $\beta\in \left[ 1,30\right] $. }}
%\label{pferro2.eps}
%\end{figure}}}%
%BeginExpansion
%EndExpansion

Note that the reentering behavior does not seem to appear at fixed densities
$\rho ^{\left( \mathit{f}\right) }\in \left( 0,2\right) $ of
\textquotedblleft $\mathit{f}$\textquotedblright --fermions, which uniquely
define at fixed $\mu _{\mathit{s}}\in \mathbb{R}$ a chemical potential $\mu
_{\mathit{f}\mathrm{,}N,\beta }$ for any $N\in \mathbb{N}$ such that
\begin{equation*}
\rho ^{\left( \mathit{f}\right) }=\omega _{N}\left( n_{x,\uparrow }^{(%
\mathit{f})}+n_{x,\downarrow }^{(\mathit{f})}\right) .
\end{equation*}%
Therefore, if the two--band model $\mathrm{H}_{N}^{(\mathit{s},\mathit{f})}$
is supposed to describe a system of fixed (fermionic) atoms (playing the
role of \textquotedblleft $\mathit{f}$\textquotedblright --fermions) and
moving fermions (\textquotedblleft $\mathit{s}$\textquotedblright \textbf{--}%
fermions), then the model does not seem -- in this situation -- to show the
reentering behavior for any choice parameters.

\begin{figure}[hbtp]
\includegraphics[angle=0,scale=1,clip=true,width=6.5cm]{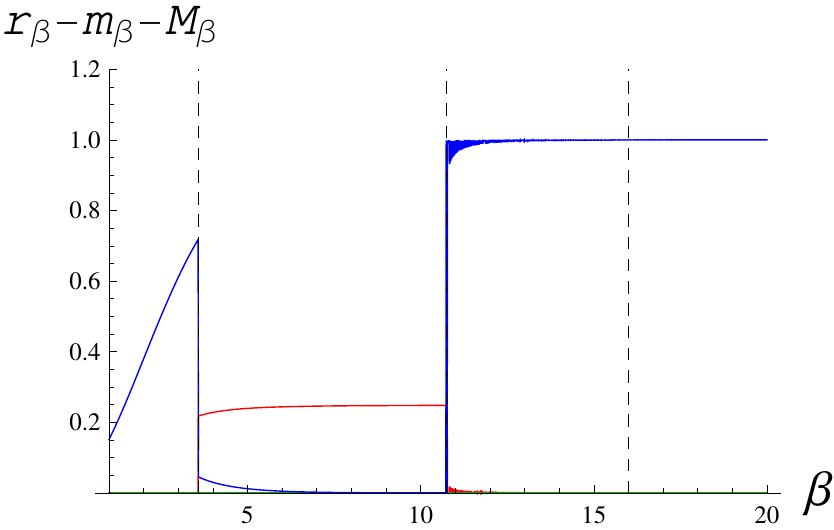}
\caption{\emph{Illustration of the Cooper pair
condensate density $\textit{r}_{\beta }$ (red), the intra-band magnetization density
$\mathrm{m}_{\beta }$ (green), and the inter-band magnetization
density $\mathrm{M}_{\beta }$ (blue) for $\gamma _{\textit{s}}=3.54$,
$\lambda _{\textit{s}}=\lambda _{\textit{f}}=0.5$, $\delta=-0.5$,
$h_{\textit{s}}=h_{\textit{f}}=\lambda =0$, $\eta=0.4$, $\rho=1$, and $\beta\in \left[ 1,20\right] $. A coexistence of phases takes place for inverse temperatures $\beta \in ( \beta _{c_{2}},\beta _{c_{3}})$ with  $\beta _{c_{2}}\simeq 10.75$ and $\beta _{c_{3}} \simeq 16$.}}
\label{ferro5.eps}
\end{figure}

\begin{figure}[hbtp]
\includegraphics[angle=0,scale=1,clip=true,width=6.5cm]{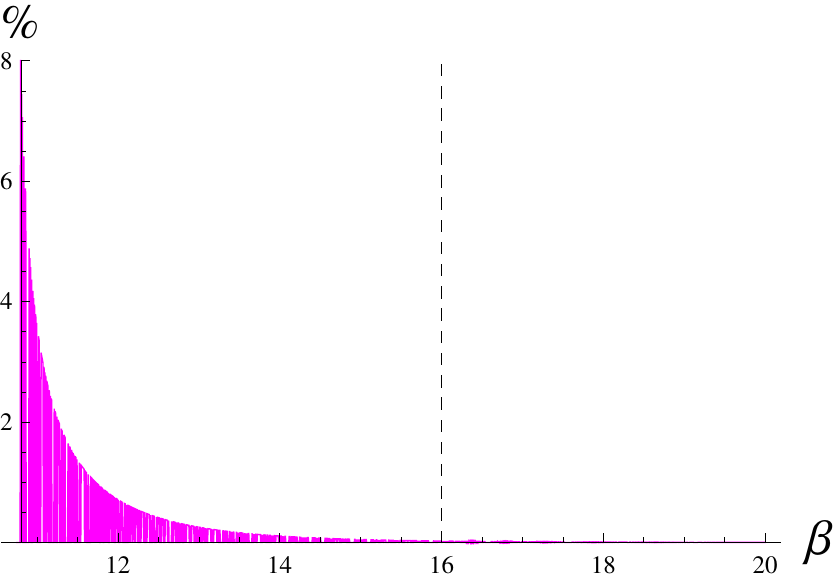}
\caption{\emph{\% of the superconducting phase for $\gamma _{\textit{s}}=3.54$,
$\lambda _{\textit{s}}=\lambda _{\textit{f}}=0.5$, $\delta=-0.5$,
$h_{\textit{s}}=h_{\textit{f}}=\lambda =0$, $\eta=0.4$, $\rho=1$, and $\beta \in (10.8, 20 )$. The black dashed line corresponds to $\beta = \beta _{c_{3}} \simeq 16$.}}
\label{ferro519.eps}
\end{figure}

At fixed total fermion density $\rho \in \left( 0,2\right) $ (cf. (\ref{mu
fixed particle density})) we have found that the reentering behavior does
not appear unless $\rho =1$. Indeed, if $\rho =1$, then -- up to numerical
aberrations we might not see -- the two--band model seems to show,
surprisingly, a reentering behavior in pretty good accordance with the
phenomenology of ferromagnetic superconductors (see Figs. \ref{ferro5.eps}
and \ref{ferro519.eps}):

\begin{itemize}
\item At all temperatures $\theta >\theta _{c_{1}}\simeq 0.28$, the system
is not superconducting.

\item At all temperatures $\theta \in (\theta _{c_{2}},\theta _{c_{1}}]$
with $\theta _{c_{2}}\simeq 0.09$, the system is superconducting.

\item At temperatures $\theta \in (\theta _{c_{3}},\theta _{c_{2}})$ with $%
\theta _{c_{3}}\simeq 0.06$, a coexistence of the non--superconducting
(which is magnetic in the presence of any non--vanishing external magnetic
field) and superconducting phases appears. This seems to be the case in real
ferromagnetic superconductors, cf. \cite[p. 263-267]{Shrivastava}.

\item At all temperatures $\theta \leq \theta _{c_{3}}$, there is a
non--superconducting -- ferromagnetic in presence of non--zero external
magnetic fields -- phase.
\end{itemize}

\noindent Note that the possible choices of parameters leading to this
peculiar behavior seems to be relatively rare.%
%TCIMACRO{%
%\TeXButton{ferro5.eps}{\begin{figure}[hbtp]
%\includegraphics[angle=0,scale=1,clip=true,width=6.5cm]{ferro5}
%\caption{\emph{Illustration of the Cooper pair
%condensate density $\textit{r}_{\beta }$ (red), the intra-band magnetization density
%$\mathrm{m}_{\beta }$ (green), and the inter-band magnetization
%density $\mathrm{M}_{\beta }$ (blue) for $\gamma _{\textit{s}}=3.54$,
%$\lambda _{\textit{s}}=\lambda _{\textit{f}}=0.5$, $\delta=-0.5$,
%$h_{\textit{s}}=h_{\textit{f}}=\lambda =0$, $\eta=0.4$, $\rho=1$, and $\beta\in \left[ 1,20\right] $. A coexistence of phases takes place for inverse temperatures $\beta \in ( \beta _{c_{2}},\beta _{c_{3}})$ with  $\beta _{c_{2}}\simeq 10.75$ and $\beta _{c_{3}} \simeq 16$.}}
%\label{ferro5.eps}
%\end{figure}}}%
%BeginExpansion
%EndExpansion
%TCIMACRO{%
%\TeXButton{ferro519.eps}{\begin{figure}[hbtp]
%\includegraphics[angle=0,scale=1,clip=true,width=6.5cm]{ferro519}
%\caption{\emph{\% of the superconducting phase for $\gamma _{\textit{s}}=3.54$,
%$\lambda _{\textit{s}}=\lambda _{\textit{f}}=0.5$, $\delta=-0.5$,
%$h_{\textit{s}}=h_{\textit{f}}=\lambda =0$, $\eta=0.4$, $\rho=1$, and $\beta \in (10.8, 20 )$. The black dashed line corresponds to $\beta = \beta _{c_{3}} \simeq 16$.}}
%\label{ferro519.eps}
%\end{figure}}}%
%BeginExpansion
%EndExpansion

At fixed chemical potentials or at fixed total fermion density per band
exactly equal one and if the parameters are chosen such that the reentering
behavior takes place, then we observe the following property of the
non--superconducting phase at low temperatures: The probability of finding
exactly one \textquotedblleft $\mathit{s}$\textquotedblright \textbf{--}%
fermion and exactly one \textquotedblleft $\mathit{f}$\textquotedblright
\textbf{--}fermion on a given lattice site goes to one as $\beta \rightarrow
\infty $. This can be interpreted as follows: At low temperatures there is
formation of pairs composed by one \textquotedblleft $\mathit{s}$%
\textquotedblright \textbf{--}fermion and one \textquotedblleft $\mathit{f}$%
\textquotedblright \textbf{--}fermion, a sort of magnetic bound state (for
instance, some kind of Kondo bound state). These pairs form in turn a Mott
phase. This mechanism of \textquotedblleft $\mathit{sf}$\textquotedblright
--pairing could be an explanation for the disappearing of the
superconducting phase. Apparently, destroying such pairs in order to set
\textquotedblleft $\mathit{s}$\textquotedblright \textbf{--}fermions free to
form Cooper pairs is energetically not favorable at low temperatures.

\section{Appendix\label{appendix}}

Thermodynamic properties of Hamiltonians of the form
\begin{equation*}
H_{N,\gamma }:=H_{N}-\frac{\gamma }{N}\sum_{x,y\in \Lambda
_{N}}a_{x,\uparrow }^{\ast }a_{x,\downarrow }^{\ast }a_{y,\downarrow
}a_{y,\uparrow }\otimes \mathbf{1}
\end{equation*}%
acting on the tensor product of Fock spaces of \textquotedblleft $\mathit{s}$%
\textquotedblright --\ and \textquotedblleft $\mathit{f}$\textquotedblright
--fermions can be analyzed by the approximating Hamiltonians%
\begin{eqnarray*}
H_{N,\gamma }\left( c\right)  &:&=H_{N} \\
&&-\frac{\gamma }{N}\sum_{x\in \Lambda _{N}}\left( \left( Nc\right)
a_{x,\uparrow }^{\ast }a_{x,\downarrow }^{\ast }+\left( N\bar{c}\right)
a_{x,\downarrow }a_{x,\uparrow }\right) \otimes \mathbf{1}
\end{eqnarray*}%
with $c\in \mathbb{C}$ conveniently chosen. This procedure is known in
mathematical physics as the \textquotedblleft approximating Hamiltonian
method\textquotedblright\ \cite{approx-hamil-method}. This was shown to be
exact for a large class of models on the level of the grand--canonical
pressure as soon as one maximizes over $c\in \mathbb{C}$ the (infinite
volume) pressure associated with $H_{N,\gamma }\left( c\right) $, see \cite%
{approx-hamil-method,BruPedra2}. The maximizers of the $c$--depending
pressures are solutions of Euler--Lagrange equations called gap equations.

Applying this method to the model $\mathrm{H}_{N}^{(\mathit{s},\mathit{f})}$
one obtains Theorem \ref{BCS theorem 1} because the approximating
Hamiltonian $\mathrm{H}_{N}^{(\mathit{s},\mathit{f})}\left( c\right) $ is a
(tensor product of the same) ($16\times 16)$--matrix which can be exactly
diagonalized. In particular, its pressure $p(c)$ can explicitly be computed
for all $c\in \mathbb{C}$. Since $\mathrm{H}_{N}^{(\mathit{s},\mathit{f})}$
is gauge invariant it suffices to restrict the variational problem to
positive real numbers $r:=|c|^{2}\geq 0$. For more details we recommend \cite%
{BruPedra1}. In case of interest, see also \cite{BruPedra2} where this
theory is developed in a much more general setting.

Proofs of (\ref{griffiths1}), (\ref{griffiths2}), (\ref{griffiths3}), (\ref%
{eq 20}), and (\ref{griffiths5}) follow from a study of equilibrium states
of the model, see \cite{BruPedra1,BruPedra2}. Heuristically, they can be
obtained by using a rather old method: Griffiths arguments which are based
on convexity properties of the pressure, see \cite[Section 8]{BruPedra1}.
The drawback of Griffiths arguments is that it requires the
differentiability of the order parameter $\mathit{r}_{\beta }$ w.r.t.
perturbations corresponding to the observable to be analyzed. The latter
represents often a difficult task and is generally even wrong at critical
points. Forgetting this problem for a moment we can compute all expectation
values. For instance, Griffiths arguments tell us that
\begin{equation*}
\underset{N\rightarrow \infty }{\lim }\left\{ N^{-1}\omega _{N}\left(
\mathfrak{c}_{0}^{\ast }\mathfrak{c}_{0}\right) \right\} =\partial _{\gamma }%
\mathrm{p|}_{\mathit{r}_{\beta }}=\mathit{r}_{\beta }
\end{equation*}%
because of the Euler--Lagrange (gap) equation (\ref{BCS gap equation}).
Similarly, $\mathit{d}_{\beta }^{\left( \mathit{s}\right) }=\partial _{\mu _{%
\mathit{s}}}\mathrm{p|}_{\mathit{r}_{\beta }}$, $\mathit{d}_{\beta }^{\left(
\mathit{f}\right) }=\partial _{\mu _{\mathit{f}}}\mathrm{p|}_{\mathit{r}%
_{\beta }}$, $\mathit{m}_{\beta }=\partial _{h_{\mathit{f}}}\mathrm{p|}_{%
\mathit{r}_{\beta }}+\partial _{h_{\mathit{s}}}\mathrm{p|}_{\mathit{r}%
_{\beta }}$ and $\mathit{M}_{\beta }=\partial _{\eta }\mathrm{p|}_{\mathit{r}%
_{\beta }}$. Computing all these derivatives by using the gap equation (\ref%
{BCS gap equation}) one obtains Equations (\ref{griffiths1}), (\ref%
{griffiths2}), (\ref{griffiths3}), (\ref{eq 20}), and (\ref{griffiths5}).

One of the main achievements of \cite{BruPedra1,BruPedra2} was to develop a
method to overcome the differentiability needed in Griffiths arguments. The
method of \cite{BruPedra1,BruPedra2} permits, moreover, to represent
equilibrium states of models in an efficient way. This makes, among other
things, the analysis of arbitrary correlation functions and the study of
equilibrium states at fixed fermion densities possible.\\[2ex]
\noindent \textit{Acknowledgments:} We thank M. Salmhofer for interesting
discussions and encouragement as well as E. Sherman for giving us
interesting references and remarks. The two first authors were supported by
the grant MTM2010-16843 of the Spanish \textquotedblleft Ministerio de
Ciencia e Innovaci{\'{o}}n\textquotedblright , the third author by a grant
of the \textquotedblleft Inneruniversit{\"{a}}re Forschungsf{\"{o}rderung}%
\textquotedblright\ of the Johannes Gutenberg University in Mainz.

\end{document}